\documentclass[10pt, prb, aps, twocolumn, showpacs, citeautoscript, floatfix, reprint, amsmath, amssymb, notitlepage, superscriptaddress]{revtex4-1}

\usepackage{graphicx}
\usepackage{stmaryrd}
\usepackage{rotating}
\usepackage{amsmath}
\usepackage{amsfonts}
\usepackage{amssymb}
\usepackage{wasysym}
\usepackage[countmax]{subfloat}
\usepackage{dcolumn} 
\usepackage{bm} 
\usepackage{color}

\usepackage{float}
\usepackage{latexsym}
\usepackage{soul}

\begin{document}

\title{Detecting the spread of valence band Wannier functions by optical sum rules}



\author{Luis F. C\'{a}rdenas-Castillo}

\affiliation{Department of Physics, PUC-Rio, 22451-900 Rio de Janeiro, Brazil}

\author{Shuai Zhang}

\affiliation{Department of Physics, PUC-Rio, 22451-900 Rio de Janeiro, Brazil}

\author{Denis Kochan}

\affiliation{Institute of Physics, Slovak Academy of Sciences, 84511 Bratislava, Slovakia}\
\affiliation{Center for Quantum Frontiers of Research and Technology (QFort), National Cheng Kung University, Tainan 70101, Taiwan}

\author{Fernando L. Freire Jr.}

\affiliation{Department of Physics, PUC-Rio, 22451-900 Rio de Janeiro, Brazil}

\author{Wei Chen}

\affiliation{Department of Physics, PUC-Rio, 22451-900 Rio de Janeiro, Brazil}

\date{\rm\today}

\begin{abstract}

The spread of valence band Wannier functions in semiconductors and insulators is a characteristic property that gives a rough estimation of how insulating is the material. We elaborate that the gauge-invariant part of the spread can be extracted experimentally from optical conductivity and absorbance, owing to their equivalence to the quantum metric of the valence band states integrated over momentum. Because the quantum metric enters the matrix element of optical conductivity, the spread of valence band Wannier functions in the gapped 3D materials can be obtained from the frequency-integration of the imaginary part of the dielectric function. We demonstrate this practically for typical semiconductors like Si and Ge, and for topological insulators like Bi$_{2}$Te$_{3}$. In 2D materials, the spread of Wannier functions in the valence bands can be obtained from the absorbance divided by frequency and then integrated over frequency. Applying this method to graphene reveals a finite spread caused by intrinsic spin-orbit coupling, which may be detected by absorbance in the microwave range. The absorbance of twisted bilayer graphene in the millimeter wave range can be used to detect the formation of the flat bands and quantify their quantum metric. Finally, we apply our method to hexagonal transition metal dichalcogenides $MX_{2}$ ($M$ = Mo, W; $X$ = S, Se, Te) and demonstrate how other effects like substrate, excitons, and higher energy bands can affect the spread of Wannier function.

\end{abstract}

\maketitle

\section{Introduction}

A recent surge in research on the quantum geometry of Bloch states has revolutionized our understanding of many physical phenomena in solids.
Focusing on fully gapped materials such as semiconductors, insulators, and superconductors, the notion of quantum geometry arises from considering the overlap between the states at neighboring momenta\cite{Provost80}, which geometrically endows the Brillouine zone (BZ), seen as compact torus $T^{D}$, with a sort of metric tensor. Many material characteristics computed in terms of  Bloch states can thus be expressed in terms of the elements of such metric tensor. For instance, in topological insulators and topological superconductors, the determinant of quantum metric is equal to the modulus of the curvature that integrates to the topological invariant\cite{vonGersdorff21_metric_curvature}, which also imposes a bound on the volume of the curved BZ\cite{Mera22}. In addition, because the quantum metric in semiconductors and insulators is also capable of providing the matrix element of optical transitions, the particular elements of quantum metric can be in principle extracted from pump-probe experiments\cite{vonGersdorff21_metric_curvature}, or from dielectric function\cite{Komissarov23}, what is further allowing conceptual generalizations of the quantum metric towards interacting systems at finite temperature\cite{Chen22_dressed_Berry_metric}. In comparison, the quantum metric of Cooper pair states also enter the optical responses of conventional superconductors in a similar way\cite{Porlles23_SC_metric}, and the metric of flat bands in the normal state has been linked to the superfluid density in the superconducting state\cite{Peotta15,Julku16,Liang17,HerzogArbeitman22,Torma22,Iskin23}.



In this paper, we focus on a particularly important ground state property of semiconductors and insulators that is directly determined by the quantum metric, namely the \textit{spread of valence band Wannier functions}. This spread measures the second cumulant of the charge distribution of the Wannier functions contributed from all the valence 
bands---a feature that has been highly exploited in first-principle calculations\cite{Marzari97,Marzari12} and the consequent Wannierization procedures. Our main goal is to make clear connections between the gauge-invariant part of the spread of Wannier function, the average of quantum metric over the BZ, and the optical matrix elements entering absorption and optical conductivity tensor of a given material, and based on them provide a feasible experimental protocol to measure the gauge-invariant part of the spread by optical or absorption means. 

For 3D materials, we show that the spread can be extracted from the frequency-integration of the imaginary part of the dielectric function and the volume of the unit cell without any other fitting parameters. Using existing experimental data of the dielectric function\cite{Madelung04,Reed99,Greenaway65}, we apply this principle and extract the absolute scale of the spread for typical semiconductors, like Si and Ge, and also for topological insulators like Bi$_{2}$Te$_{3}$. This demonstrates the ubiquity of our method for applications in any 3D semiconducting and insulating material. 

For 2D materials, we reveal that the spread is given by the absorbance divided by frequency followed by the integration of such ratio over frequency, which again requires no fitting parameters other than the unit cell area. We choose the following three 2D materials to demonstrate the ubiquity of our method, despite their frequency ranges differ by several orders of magnitude. For graphene, we uncover that the intrinsic spin-orbit coupling (ISOC) causes a finite spread\cite{Kochan17}, and moreover it cures the divergence of quantum metric at the Dirac points. The microwave absorbance measured in the sub-Kelvin region is proposed to quantify the magnitude of ISOC in graphene. For twisted bilayer graphene (TBG), we use a recently proposed tight-binding model\cite{Bennett23} to elaborate how absorbance depends on the twist angle and how that can be used to experimentally detect the formation of the flat bands and quantify their quantum metric. Finally, we turn to tight-binding models of hexagonal transition metal dichalcogenides (TMDs) to estimate their spread, and elaborate how the latter being effected by substrates, excitons, and higher energy orbitals. We particularly compare model calculations in WS$_{2}$ with the experimental data for WS$_{2}$ deposited on fused silica, confirming that the proposed experimental protocol can be performed in realistic measurement.


\section{Quantum metric, Optical conductivity, and spread of valence band Wannier functions}

\subsection{Relating the optical conductivity to quantum geometry \label{sec:opacity_to_quantum_geometry}}

We consider fully gapped semiconductors and insulators and work in the SI unit. The momentum is denoted by ${\bf k}$ and has units of $[{\bf k}]=$kg\,m/s. We use the index $n$ for valence bands, $m$ for conduction bands, and $\ell$ for all the bands at momentum ${\bf k}$. Moreover, we denote by $\langle{\bf r}|\ell\rangle=\ell({\bf r})=e^{-i{\bf k\cdot r}/\hbar}\psi_{\ell}^{\bf k}({\bf r})$ the periodic part of the single-particle Bloch state $\psi_{\ell}^{\bf k}({\bf r})$ that satisfies $\ell({\bf r})=\ell({\bf r+R})$. The ${\bf r}$ and ${\bf R}$ are, 
respectively, the position and Bravais lattice vectors. The corresponding Wannier state $|{\bf R}\ell\rangle$ is given by
\begin{eqnarray}
&&|\ell\rangle=\sum_{{\bf R}}e^{-i {\bf k}\cdot({\hat{\bf r}}-{\bf R})/\hbar}|{\bf R}\ell\rangle,
\nonumber \\
&&|{\bf R} \ell\rangle=\sum_{\bf k}e^{i {\bf k}\cdot({\hat{\bf r}}-{\bf R})/\hbar}|\ell\rangle,\;\;\;\;\;
\label{Wannier_basis}
\end{eqnarray}
while $\langle {\bf r}|{\bf R} \ell\rangle=W_{\ell}({\bf r}-{\bf R})$ stands for the conventional Wannier function of charge carrier in the $\ell$th-band located around the unit cell at ${\bf R}$.
Suppose the system has a gap and occupies $N_{-}$ valence bands (including spin or any other quantum number), then the fully antisymmetric many-body valence band Bloch state is given by
\begin{eqnarray}
|u^{\rm val}({\bf k})\rangle=\frac{1}{\sqrt{N_{-}!}}\epsilon^{n_{1}n_{2}...n_{N-}}|n_{1}\rangle|n_{2}\rangle...|n_{N_{-}}\rangle.\;\;\;
\label{psi_val}
\end{eqnarray}
The main ingredient in our formalism is the quantum metric defined from the overlap of neighboring valence band states in momentum space\cite{Provost80} 
\begin{eqnarray}
|\langle u^{\rm val}({\bf k})|u^{\rm val}({\bf k+\delta k})\rangle|=1-\frac{1}{2}g_{\mu\nu}({\bf k})\delta k^{\mu}\delta k^{\nu},
\label{uval_gmunu}
\end{eqnarray}
which amounts to the expression\cite{Matsuura10,vonGersdorff21_metric_curvature} 
\begin{eqnarray}
&&g_{\mu\nu}({\bf k})=\frac{1}{2}\langle \partial_{\mu}u^{\rm val}|\partial_{\nu}u^{\rm val}\rangle+\frac{1}{2}\langle \partial_{\nu}u^{\rm val}|\partial_{\mu}u^{\rm val}\rangle
\nonumber \\
&&-\langle \partial_{\mu}u^{\rm val}|u^{\rm val}\rangle \langle u^{\rm val}|\partial_{\nu}u^{\rm val}\rangle
\nonumber \\
&&=
\frac{1}{2}\sum_{nm}\left[\langle \partial_{\mu}n|m\rangle\langle m|\partial_{\nu}n\rangle+\langle \partial_{\nu}n|m\rangle\langle m|\partial_{\mu}n\rangle\right],
\label{gmunu_T0}
\end{eqnarray}
with $\partial_{\mu}\equiv\partial/\partial k^{\mu}$. 
The key quantum geometrical quantity that is related to the spread of Wannier functions is the 
momentum integral\cite{Souza08,Marzari97,Marzari12} of $g_{\mu\nu}({\bf k})$ 
\begin{eqnarray}
&&{\cal G}_{\mu\nu}=\int\frac{d^{D}{\bf k}}{(2\pi)^{D}}g_{\mu\nu}({\bf k}),
\label{Gmunu_definition}
\end{eqnarray}
which we call the fidelity number\cite{deSousa23_fidelity_marker}. 

To link these geometrical quantities to experimentally measurable data, it is practical to introduce a quantum metric spectral function $g_{\mu\mu}^{d}({\bf k},\omega)$ defined as the real part of the longitudinal optical conductivity at momentum ${\bf k}$, frequency $\omega$ and polarization ${\hat{\boldsymbol\mu}}$\cite{Mahan00,Chen22_dressed_Berry_metric,deSousa23_fidelity_marker}
\begin{eqnarray}
&&\sigma_{\mu\mu}({\bf k},\omega)=\sum_{\ell<\ell '}
\frac{\pi}{V_{\rm cell}\,\hbar\omega}
\langle\ell|{\hat j}_{\mu}|{\ell '} \rangle\langle{\ell '}|{\hat j}_{\mu}|\ell\rangle
\nonumber \\
&&\times\left[f(\varepsilon_{\ell}^{\bf k})-f(\varepsilon_{\ell '}^{\bf k})\right]\delta\left(\omega+\frac{\varepsilon_{\ell}^{\bf k}}{\hbar}-\frac{\varepsilon_{\ell '}^{\bf k}}{\hbar}\right)
\nonumber \\
&&=\frac{\pi e^{2}}{V_{\rm cell}}\hbar\omega\,g_{\mu\mu}^{d}({\bf k},\omega).
\label{sigmaw_gmumu}
\end{eqnarray}
where $f(\varepsilon_{\ell}^{\bf k})$ is the Fermi distribution function at the eigenenergy $\varepsilon_{\ell}^{\bf k}$, $V_{\rm cell}$ is the volume of the unit cell, and $\hat{j}_{\mu}$ represents the $\mu$-th component of the current operator. The superscript $d$ stands as an acronym for "dressed" since the formalism based on optical conductivity can incorporate finite temperature and many-body interactions.
The frequency integral of $g_{\mu\mu}^{d}({\bf k},\omega)$ gives the dressed quantum metric $g_{\mu\mu}^{d}({\bf k})=\int_{0}^{\infty}d\omega\,g_{\mu\mu}^{d}({\bf k},\omega)$, as has been pointed out previously\cite{Souza00,Resta02,Resta11,Chen22_dressed_Berry_metric,Kashihara22}. In the zero temperature and noninteracting limit $\lim_{T\rightarrow 0}g_{\mu\nu}^{d}({\bf k})=g_{\mu\nu}({\bf k})$, the frequency integration recovers to the expression given by Eq.~(\ref{gmunu_T0}).

The optical conductivity measured in a real space corresponds to the momentum integration of $\sigma_{\mu\mu}({\bf k},\omega)$ over the BZ of volume $V_{BZ}=(2\pi)^{D}/V_{\rm cell}$
\begin{eqnarray}
&&\sigma_{\mu\mu}(\omega)=\int\frac{d^{D}{\bf k}}{\hbar^{D}V_{BZ}}\,\sigma_{\mu\mu}({\bf k},\omega)
\nonumber \\
&&=\frac{\pi e^{2}}{\hbar^{D-1}}\,\omega\int\frac{d^{D}{\bf k}}{(2\pi)^{D}}\,g_{\mu\mu}^{d}({\bf k},\omega)
\equiv\frac{\pi e^{2}}{\hbar^{D-1}}\,\omega\,{\cal G}_{\mu\mu}^{d}(\omega),
\label{fidelity_number_spec_fn}
\end{eqnarray}
leading to the dressed fidelity-number spectral function ${\cal G}_{\mu\mu}^{d}(\omega)$ at polarization ${\hat{\boldsymbol\mu}}$, whose frequency integral gives the dressed fidelity number
\begin{eqnarray}
{\cal G}_{\mu\mu}^{d}=\int_{0}^{\infty}d\omega\,{\cal G}_{\mu\mu}^{d}(\omega),
\label{fidelity_number_spec_fn_int}
\end{eqnarray}
which is the key ingredient to link the quantum geometry to the optical absorption power, as we shall see below for 3D and 2D materials. call this quantity ${\cal G}_{\mu\mu}^{d}$ the fidelity number because it is dimensionless in 2D. Although in 3D it has the unit of momentum, one can trivially multiply it by $V_{\rm cell}^{1/3}/\hbar$ to make it dimensionless, and hence we stick with this nomenclature. 


\subsection{Relating quantum geometry to the spread}

We proceed by elaborating on a connection between the fidelity number and the spread of valence band Wannier functions defined by\cite{Marzari97,Marzari12}
\begin{eqnarray}
&&\Omega=\sum_{n}\left[\langle r^{2}\rangle_{n}-{\hat{\bf r}}_{n}^{2}\right]
\nonumber \\
&&=\sum_{n}\left[\langle{\bf 0}n|r^{2}|{\bf 0}n\rangle-\langle{\bf 0}n|{\bf r}|{\bf 0}n\rangle^{2}\right]=\Omega_{I}+\tilde{\Omega},
\label{Omega_original}
\end{eqnarray}
This spread represents the variance of charge distribution associated with the valence band Wannier states of the given material.
As the single-particle Bloch states are unique up to a gauge-transformation, 
$\psi_{\ell}^{\bf k}({\bf r}) \mapsto e^{i\varphi_\ell(\bf{k})}\,\psi_{\ell}^{\bf k}({\bf r})$, involving a momentum-dependent phase 
$\varphi_\ell(\mathbf{k})$, $\Omega$ itself can be separated into the gauge invariant and gauge-dependent parts
\begin{eqnarray}
&&\Omega_{I}=\sum_{n}\left[\langle{\bf 0}n|r^{2}|{\bf 0}n\rangle-\sum_{{\bf R}n'}|\langle{\bf R}n'|{\bf r}|{\bf 0}n\rangle|^{2}\right],
\nonumber \\
&&\tilde{\Omega}=\sum_{n}\sum_{{\bf R}n'\neq{\bf 0}n}|\langle{\bf R}n'|{\bf r}|{\bf 0}n\rangle|^{2},
\label{Omega_gauge_inv}
\end{eqnarray}
that are central to the concept of Wannierization and maximally localized Wannier orbitals\cite{Marzari97,Marzari12}. 
Contrary to the Wannierization schemes where the summations over $n$ and $n'$ are usually limited to a few bands near the Fermi level or the gap, in Eq.~(\ref{Omega_gauge_inv}) we account for the summations over the whole valence band manifold $\left\{n,n'\right\}\in v$. For these reasons $\Omega_{I}$ is a fully gauge-invariant quantity characterizing the ground state. Moreover, $\Omega$ represents the second cumulant in the theory of charge polarization, as can be understood by comparing it with the definition of the charge polarization
\begin{eqnarray}
{\bf P}=\sum_{n}\langle{\bf 0}n|{\bf r}|{\bf 0}n\rangle,
\end{eqnarray}
which measures the first cumulant of the charge distribution of valence band Wannier states. Just like the first cumulant ${\bf P}$, the second cumulant in Eq.~(\ref{Omega_original}) is also gauge-dependent, but the gauge-invariant part $\Omega_{I}$ in Eq.~(\ref{Omega_gauge_inv}) should be physically measurable. 


The main goal of the present work is to propose an experimental protocol to determine $\Omega_{I}$ based on its equivalence with the fidelity number ${\cal G}_{\mu\mu}$, and the connection of the latter with the longitudinal optical conductivity $\sigma_{\mu\mu}(\omega)$ according to Eqs.(\ref{fidelity_number_spec_fn}) and (\ref{fidelity_number_spec_fn_int}).
This equivalence can be seen by considering the identities 
\begin{align}
\langle r^{2}\rangle_{n} &=\frac{V_{\rm cell}}{\hbar^{D-2}}\int\frac{d^{D}{\bf k}}{(2\pi)^{D}}\sum_{\mu}\langle\partial_{\mu}n|\partial_{\mu}n\rangle,
\nonumber \\
\langle{\bf R}n'|\hat{r}_{\mu}|{\bf 0}n\rangle &=\frac{V_{\rm cell}}{\hbar^{D-1}}\int\frac{d^{D}{\bf k}}{(2\pi)^{D}}\langle n'|i\partial_{\mu}|n\rangle e^{i{\bf k\cdot R}/\hbar},
\end{align}
from which $\Omega_{I}$ can be written as\cite{Souza08,Marzari97,Marzari12} 
\begin{eqnarray}
&&\Omega_{I}=\frac{V_{\rm cell}}{\hbar^{D-2}}\int\frac{d^{D}{\bf k}}{(2\pi)^{D}}
\nonumber \\
&&\times\sum_{\mu}\sum_{n}
\left[\langle\partial_{\mu}n|\partial_{\mu}n\rangle-\sum_{n'}\langle\partial_{\mu}n|n'\rangle\langle n'|\partial_{\mu}n\rangle\right]
\nonumber \\
&&=\frac{V_{\rm cell}}{\hbar^{D-2}}\,{\rm Tr}\,{\cal G}_{\mu\nu}=\frac{V_{\rm cell}}{\hbar^{D-2}}\sum_{\mu}{\cal G}_{\mu\mu}\,,
\label{OmegaI_trace_Gmunu}
\end{eqnarray}
indicating that the gauge-invariant part of the spread of valence band Wannier states is equivalent to the trace of fidelity number (tensor). Thus the quantum metric formalism developed in Sec.~\ref{sec:opacity_to_quantum_geometry} linking the fidelity number to optical conductivity can be particularly useful to extract $\Omega_{I}$ experimentally, as we shall see in the following sections devoted to particular 3D and 2D systems.

\section{Applications to 3D systems}

\subsection{Extracting the spread of Wannier functions in 3D systems from dielectric function}

As shown explicitly below for 3D materials, the frequency-integration of the imaginary part of dielectric function is directly proportional to the fidelity number ${\cal G}_{\mu\mu}$ and hence to the spread $\Omega_{I}$. 
Considering a propagation of oscillating electric field that varies on a spatial scale far above the lattice constant $a$, resulting in a relation between the complex dielectric function and the optical conductivity\cite{Ashcroft76}
\begin{eqnarray}
&&\varepsilon_{\mu\nu}(\omega)
=1+i\frac{\tilde{\sigma}_{\mu\nu}(\omega)}{\varepsilon_{0}\omega}\,,
\end{eqnarray}
where $\varepsilon_{0}$ is the vacuum permittivity, and $\tilde{\sigma}_{\mu\nu}(\omega)$ is the complex conductivity whose real part in the longitudinal direction is the $\sigma_{\mu\mu}(\omega)$ in Eq.~(\ref{sigmaw_gmumu}). Thus the imaginary part of dielectric function is given by the real part of the optical conductivity, and hence the diagonal elements of ${\rm Im}[\varepsilon_{\mu\mu}(\omega)]$ are related to the fidelity number spectral function by
\begin{eqnarray}
{\rm Im}[\varepsilon_{\mu\mu}(\omega)]=\frac{1}{\varepsilon_{0}\omega}\sigma_{\mu\mu}(\omega)=\frac{\pi e^{2}}{\varepsilon_{0}\hbar^{2}}{\cal G}_{\mu\mu}^{d}(\omega)\,.
\end{eqnarray} 
As a result, the dressed fidelity number and the spread of valence band Wannier function are directly determined by the frequency/energy integrals of ${\rm Im}[\varepsilon_{\mu\mu}(\omega)]$
\begin{eqnarray}
&&{\cal G}_{\mu\mu}^{d}=\frac{\varepsilon_{0}\hbar}{\pi e^{2}}\int_{0}^{\infty}d(\hbar\omega)\,{\rm Im}[\varepsilon_{\mu\mu}(\omega)],
\label{OmegaI_Imepsilon0} \\
&&\Omega_{I}=\lim_{T\rightarrow 0}\frac{V_{\rm cell}\varepsilon_{0}}{\pi e^{2}}\sum_{\mu}\int_{0}^{\infty}d(\hbar\omega)\,{\rm Im}[\varepsilon_{\mu\mu}(\omega)].
\label{OmegaI_Imepsilon}
\end{eqnarray}
Equation (\ref{OmegaI_Imepsilon}) is the central result of our theory for 3D systems, which serves as a concrete experimental protocol to measure $\Omega_{I}$.



\subsection{Applications to common semiconductors}

The experimental data of the frequency-dependence of the dielectric function are available for a wide variety of semiconductors. There are, however, several issues when extracting $\Omega_{I}$ from the experimental data, which we now address. 

First, most experiments are performed using an unpolarized light 
on polycrystalline samples, what is equivalent of shining a randomly polarized light on a single crystal, therefore we consider a random electric field 
\begin{eqnarray}
{\bf E}=E_{0}(\sin\theta\cos\phi,\sin\theta\sin\phi,\cos\theta)\cos{\omega t},
\end{eqnarray}
evenly distributed over the solid angle $\sin\theta d\theta d\phi$. Writing the induced electric current at frequency $\omega$ as $j_{\mu}=\sigma_{\mu\nu}E_{\nu}$, the product of electric current and electric field gives the time-averaged absorption power density
$ W_{a}(\omega,\theta,\phi)=\sum_{\mu} \langle j_{\mu}(\omega,t)E_{\mu}(\omega,t)\rangle_t=\sum_{\mu}j_{\mu}(\omega)E_{\mu}(\omega)/2=\sum_{\mu\nu}\sigma_{\mu\nu}(\omega)E_{\mu}(\omega)E_{\nu}(\omega)/2$ at frequency $\omega$ due to incident light 
from the solid angle $\sin\theta d\theta d\phi$. Here the factor of $1/2$ comes from the time-averaging of the oscillating factor
 $\langle\cos^{2}\omega t\rangle_{t}=1/2$ as both the current and field oscillate harmonically with a frequency $\omega$. 
As the one measures the absorption power averaged over the solid angle 
\begin{eqnarray}
&&\overline{W}_{a}(\omega)=\frac{1}{4\pi}\int_{0}^{2\pi}d\phi\int_{0}^{\pi}\sin\theta d\theta\,\frac{1}{2}\sum_{\mu}j_{\mu}(\omega)E_{\mu}(\omega)
\nonumber \\
&&=\frac{1}{2}\left(\frac{1}{3}\sum_{\mu}\sigma_{\mu\mu}(\omega)\right)E_{0}^{2}
\equiv\frac{1}{2}\overline{\sigma}(\omega)E_{0}^{2}.
\end{eqnarray}
the experiment has an access to directionally averaged optical conductivity $\overline{\sigma}(\omega)=\sum_{\mu}\sigma_{\mu\mu}(\omega)/3$ and consequently the directionally averaged dielectric function ${\rm Im}[\overline{\varepsilon}(\omega)]=\sum_{\mu}{\rm Im}[\varepsilon_{\mu\mu}(\omega)]/3$.
Because the $\Omega_{I}$ in Eq.~(\ref{OmegaI_Imepsilon}) requires the summation over three crystalline directions, we should multiply the experimental dielectric function by a factor of three $\sum_{\mu}{\rm Im}[\varepsilon_{\mu\mu}(\omega)]=3{\rm Im}[\overline{\varepsilon}(\omega)]$.

Second, measuring the optical absorption in the gapped materials would include apart from the valence-to-conduction band electron transitions also excitonic effects. We suggest that when integrating experimental data over frequency one should exclude poisoning due to excitonic peaks that emerge at 
frequencies inside the material band gap. The reason for this exclusion will be elaborated in Sec.~\ref{sec:WS2_silica}.

Third, the dielectric function is usually measured as a function of energy in units of eV instead of $\omega$. Thus we use the conversion 
\begin{eqnarray}
&&\int_{0}^{\infty}d(\hbar\omega)\,{\rm Im}[\overline{\varepsilon}(\omega)]
=\int_{0}^{\infty}d({\rm eV})\,{\rm Im}[\overline{\varepsilon}({\rm eV})]
\nonumber \\
&&=\nu\times {\rm eV}.
\label{nu_3D_definition}
\end{eqnarray}
i.e., the frequency-integration of ${\rm Im}[\overline{\varepsilon}({\rm eV})]$ over energy is a dimensionless number $\nu$ times the unit of eV. Putting $\nu$ and the factor of 3 as discussed above into Eq.~(\ref{OmegaI_Imepsilon}), we get 
\begin{eqnarray}
&&\Omega_{I}=\lim_{T\rightarrow 0}\frac{V_{\rm cell}\varepsilon_{0}}{\pi e^{2}}\,{\rm eV}\times 3\nu
\nonumber \\
&&\;\;\;\;\;=\lim_{T\rightarrow 0}\frac{V_{\rm cell}}{\AA}\times 1.7591\times 10^{-3}\times 3\nu,
\nonumber \\
&&{\rm Tr}{\cal G}_{\mu\nu}=\lim_{T\rightarrow 0}\frac{\hbar}{\AA}\times 1.7591\times 10^{-3}\times 3\nu,
\label{spread_3D_general_formula}
\end{eqnarray}
which serve as very simple formulas to extract $\Omega_{I}$ and ${\rm Tr}\,{\cal G}_{\mu\nu}$ experimentally. 
Furthermore, the band gap of common semiconductors is much higher than room temperature, so the $T\rightarrow 0$ limit in Eq.~(\ref{spread_3D_general_formula}) is safely achieved. 

Finally, by combining the spread $\Omega_{I}$ and $V_{\rm cell}$, one may calculate the dimensionless ratio 
$\Omega_{I}^{3/2}/V_{\rm cell}$ and interpret it as a figure of merit 
of how insulating is the given material. The smaller is this ratio, the more insulating is the material. 
In Table \ref{tab:Si_Ge_Bi2Te3_data}, we use the corresponding experimental data from Refs. \onlinecite{Reed99}, \onlinecite{Madelung04} 
and \onlinecite{Greenaway65}  for the dielectric functions of Si, Ge (semiconductors) and Bi$_{2}$Te$_{3}$ (topological insulator) and extract $\Omega_{I}$ and $\Omega_{I}^{3/2}/V_{\rm cell}$, which reveal a much more extended Wannier function in Bi$_{2}$Te$_{3}$ than in Si and Ge.

\begin{table}[ht]
  \begin{center}
    \caption{Volume of the unit cell $V_{\rm cell}$, frequency-integration $\nu$ of the imaginary part of dielectric function, trace of the fidelity number ${\rm Tr}{\cal G}_{\mu\nu}$, spread of the valence band Wannier functions $\Omega_{I}$, and the dimensionless ratio $\Omega_{I}^{3/2}/V_{\rm cell}$ extracted from experimental data for semiconductors Si and Ge, and topological insulator Bi$_{2}$Te$_{3}$.}
    \label{tab:Si_Ge_Bi2Te3_data}
    \begin{tabular}{ c c c c c c }
    \hline
    {\rm Mat}. & $V_{\rm cell}(\AA^{3})$ & $\nu$ & ${\rm Tr}{\cal G}_{\mu\nu}(\hbar/\AA)$ & $\Omega_{I}(\AA^{2})$ & $\Omega_{I}^{3/2}/V_{\rm cell}$ \\ \hline
    Si & 160.1 & 80.6 & 0.425 & 68.1 & 3.51 \\ 
    Ge & 181.3 & 87.0 & 0.459 & 83.24 & 4.19 \\ 
    Bi$_{2}$Te$_{3}$ & 545.3 & 141.6 & 0.747 & 407.48 & 15.08 \\ 
    \hline
  \end{tabular}
  \end{center}
\end{table}

\section{Applications to 2D systems}

\subsection{Detecting the spread of Wannier functions in 2D systems by absorbance}

We now turn to 2D materials, which has been partly discussed previously\cite{deSousa23_graphene_opacity}. Consider a 2D material subjected to a polarized oscillating field $E_{\mu}(\omega,t)=E_{0}\cos\omega t$ that consequently induces a current $j_{\mu}(\omega,t)=\sigma_{\mu\mu}(\omega)E_{0}\cos\omega t$. The optical absorption power density at frequency $\omega$ and polarization $\mu$ is 
\begin{eqnarray}
&&W_{a}^{\mu}(\omega)=\langle j_{\mu}(\omega,t)E_{\mu}(\omega,t)\rangle_{t}=\frac{1}{2}\sigma_{\mu\mu}(\omega)E_{0}^{2}
\nonumber \\
&&=\frac{\pi e^{2}}{2\hbar}\,E_{0}^{2}\omega\,{\cal G}_{\mu\mu}^{d}(\omega).
\label{absorption_power_global}
\end{eqnarray}
Given the incident power of the light per unit area $W_{i}=c\varepsilon_{0}E_{0}^{2}/2$, the absorbance or opacity under polarization $\mu$ and frequency $\omega$ is then\cite{Nair08} 
\begin{eqnarray}
{\cal O}_{\mu}(\omega)=\frac{W_{a}(\omega)}{W_{i}}=4\pi^{2}\alpha\omega\,{\cal G}_{\mu\mu}^{d}(\omega)|_{2D},
\label{opacity_wGw}
\end{eqnarray}
where $\alpha=e^{2}/4\pi\varepsilon_{0}\hbar c\approx 1/137$ is the fine-structure constant. From this relation, one sees that ${\cal G}_{\mu\mu}^{d}(\omega)$ can be simply extracted experimentally from the opacity by
\begin{eqnarray}
{\cal G}_{\mu\mu}^{d}(\omega)|_{2D}=\frac{1}{4\pi\omega}\left[\frac{{\cal O}_{\mu}(\omega)}{\pi\alpha}\right],
\label{fidelity_number_opacity}
\end{eqnarray}
Following the discussion in Sec.~\ref{sec:opacity_to_quantum_geometry},
in the zero temperature and noninteracting limit $\lim_{T\rightarrow 0}{\cal G}_{\mu\nu}^{d}(\omega)={\cal G}_{\mu\nu}(\omega)$, the gauge-invariant spread is simply given by
\begin{eqnarray}
&&\Omega_{I}=A_{\rm cell}{\rm Tr}\,{\cal G}_{\mu\nu}=A_{\rm cell}\int_{0}^{\infty}d\omega\,{\rm Tr}\,{\cal G}_{\mu\nu}(\omega)
\nonumber \\
&&=\lim_{T\rightarrow 0}A_{\rm cell}\sum_{\mu}\int_{0}^{\infty}d\omega\,\frac{1}{4\pi\omega}\left[\frac{{\cal O}_{\mu}(\omega)}{\pi\alpha}\right].
\label{OmegaI_opacity}
\end{eqnarray}
and hence the ratio $\Omega_{I}/A_{\rm cell}={\rm Tr}\,{\cal G}_{\mu\nu}$ that characterizes how insulating is the material is simply given by the trace of fidelity number.


If the opacity measurement is done using unpolarized light or on a polycrystalline sample, then the average absorption power is 
\begin{eqnarray}
&&\overline{W}_{a}=\int_{0}^{2\pi}\frac{d\phi}{2\pi}W_{a}(\phi)
=\frac{1}{2}\left(\frac{1}{2}\sigma_{xx}(\omega)+\frac{1}{2}\sigma_{yy}(\omega)\right)E_{0}^{2}
\nonumber \\
&&\equiv \frac{1}{2}\overline{\sigma}(\omega)E_{0}^{2}.
\end{eqnarray}
Therefore the summation over polarization $\sum_{\mu}{\cal O}_{\mu}(\omega)=2\overline{\cal O}(\omega)$ in Eq.~(\ref{OmegaI_opacity}) just gives a factor of 2, i.e., 
\begin{eqnarray}
&&\Omega_{I}=\lim_{T\rightarrow 0}A_{\rm cell}\int_{0}^{\infty}d\omega\,\frac{1}{2\pi\omega}\left[\frac{\overline{\cal O}(\omega)}{\pi\alpha}\right]
\nonumber \\
&&=A_{\rm cell}{\rm Tr}\,{\cal G}_{\mu\nu}.
\label{OmegaI_opacity}
\end{eqnarray}  
This formula provides a concrete experimental protocol to measure the gauge-invariant spread of Wannier orbitals and the trace of fidelity number, and enables their comparison with those obtained from first-principle or tight-binding model calculations. In the following sections, we use monolayer graphene, TBG, and TMDs to demonstrate the feasibility of this protocol, and give an estimation to the absolute scale of the spread.

\section{Graphene with ISOC \label{sec:graphene_ISOC}}

\subsection{Wrapping number in graphene with ISOC}

\begin{figure}[ht]
\begin{center}
\includegraphics[clip=true,width=0.99\columnwidth]{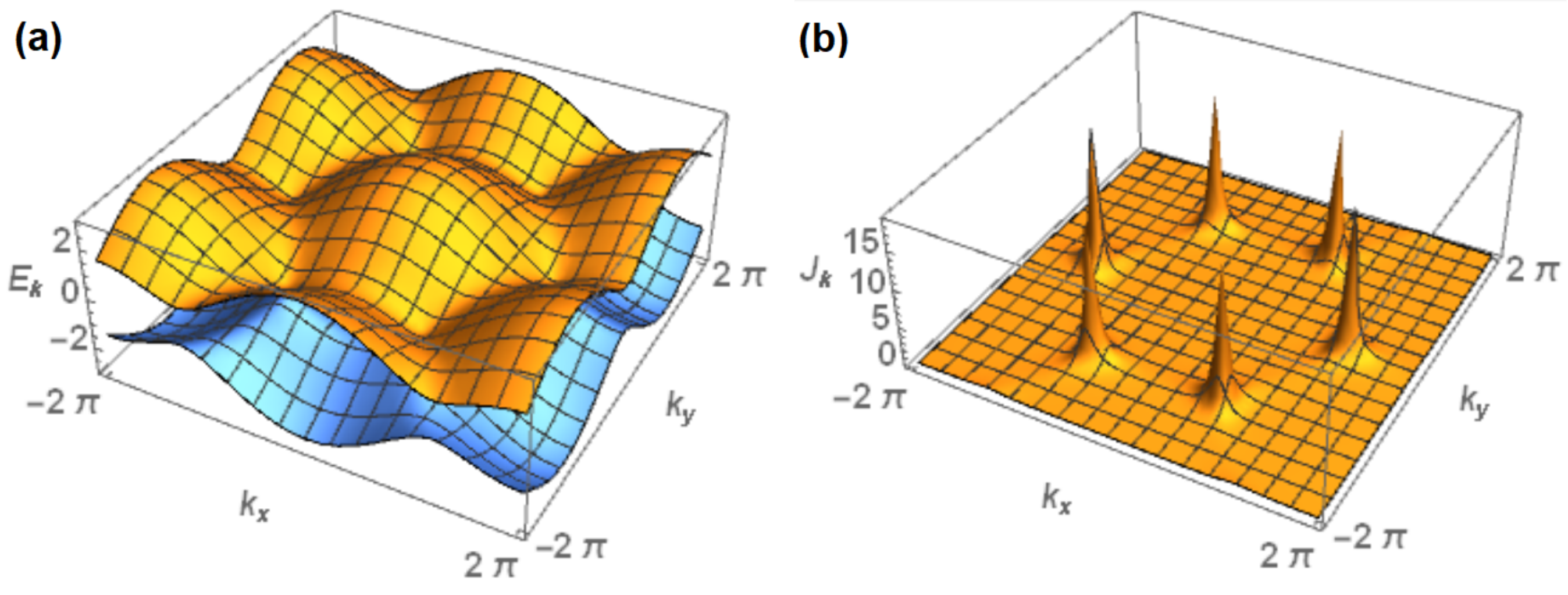}
\caption{(a) The energy dispersion and (b) the corresponding Jacobian of the $T^{2}\rightarrow S^{2}$ map for graphene with ISOC. We choose an exaggerated ISOC value $\lambda_{I}/t=0.2$ to visually demonstrate the main features. } 
\label{fig:graphene_Ek_Jk}
\end{center}
\end{figure}

Our first example is the lattice model of graphene with nearest-neighbor hopping and ISOC. Denoting the lattice constant on the honeycomb lattice to be the unit of distance $a_{L}=1$ (the carbon-carbon distance is $a_{L}/\sqrt{3}$), we define\cite{Kochan17}
\begin{eqnarray}
&&{\bf R}_{1}=(1,0)a_{L},\;\;\;{\bf R}_{2}=\left(-\frac{1}{2},\frac{\sqrt{3}}{2}\right)a_{L},
\nonumber \\
&&{\bf R}_{3}=\left(-\frac{1}{2},-\frac{\sqrt{3}}{2}\right)a_{L}.
\end{eqnarray}
The lattice model is described by the second-quantized Hamiltonian ${\cal H}=\sum_{\bf q}\psi_{\bf q}^{\dag}H({\bf q})\psi_{\bf q}$ with the basis $\psi_{\bf q}^{\dag}=(c_{A{\bf q}\uparrow}^{\dag},c_{B{\bf q}\uparrow}^{\dag},c_{A{\bf q}\downarrow}^{\dag},c_{B{\bf q}\downarrow}^{\dag})$, where $c_{X{\bf q}\sigma}^{\dag}$ is the electron creation operator at momentum ${\bf q}$ (with respect to the center of BZ), sublattice $X=\left\{A,B\right\}$, and spin 
$\sigma=\left\{\uparrow,\downarrow\right\}$. The $4\times 4$ Hamiltonian takes the form\cite{Kochan17}
\begin{eqnarray}
&&H({\bf q})=\left(\begin{array}{cccc}
\lambda_{I}f_{I} & -tf_{orb} & 0 & 0 \\
-tf_{orb}^{\ast} & -\lambda_{I}f_{I} & 0 & 0 \\
0 & 0 & -\lambda_{I}f_{I} & -tf_{orb} \\
0 & 0 & -tf_{orb}^{\ast} & \lambda_{I}f_{I}
\end{array}\right),
\nonumber \\
&&f_{orb}({\bf q})=1+e^{i{\bf q\cdot R}_{2}/\hbar}+e^{-i{\bf q\cdot R}_{3}/\hbar},
\nonumber \\
&&f_{I}({\bf q})=-\frac{2}{3\sqrt{3}}\sum_{i=1}^{3}\sin{\bf q\cdot R}_{i}/\hbar,
\label{HK_whole_BZ}
\end{eqnarray}
where $t$ and $\lambda_{I}$ are the magnitudes of nearest-neighbor orbital hopping and ISOC, respectively. The Dirac points in this model are 
located at 
\begin{eqnarray}
\frac{{\bf K}}{\hbar/a_{L}}=\left(\frac{4\pi}{3},0\right),\;\;\;
\frac{{\bf K'}}{\hbar/a_{L}}=\left(-\frac{4\pi}{3},0\right),
\label{KKp_points}
\end{eqnarray}
which are of particular interest for the optical responses, as we shall see later. The energy dispersion of this model is shown in Fig.~\ref{fig:graphene_Ek_Jk}(a).

We notice that the Hamiltonian in Eq.~(\ref{HK_whole_BZ}) is a Dirac Hamiltonian with the following representations for the $4\times 4$ Dirac matrices (see Eq.~(B.2) of Ref.~\onlinecite{Ryu10})
\begin{eqnarray}
\Gamma_{1\sim 5}=({\hat s}_{1}\otimes{\hat\sigma}_{3},{\hat s}_{2}\otimes{\hat\sigma}_{3},{\hat s}_{0}\otimes{\hat\sigma}_{1},{\hat s}_{0}\otimes{\hat\sigma}_{2},{\hat s}_{3}\otimes{\hat\sigma}_{3}).\;\;\;
\end{eqnarray}
where ${\hat s}_{\alpha}$ and ${\hat \sigma}_{\alpha}$ are the Pauli matrices in the spin space $\left\{\uparrow,\downarrow\right\}$ and the sublattice space $\left\{A,B\right\}$, respectively. In terms of the $\Gamma$-matrices, the Hamiltonian in Eq.~(\ref{HK_whole_BZ}) is expanded by $(\Gamma_{3},\Gamma_{4},\Gamma_{5})$
\begin{eqnarray}
&&H({\bf q})={\bf d}\cdot{\boldsymbol\Gamma}=d_{3}\Gamma_{3}+d_{4}\Gamma_{4}+d_{5}\Gamma_{5},
\nonumber \\
&&\left(d_{3},d_{4}\right)=-t\,\left({\rm Re},{\rm Im}\right)f_{orb}({\bf q}),\;\;\;d_{5}=\lambda_{I}f_{I}({\bf q}).
\label{Hk4by4_Gamma345}
\end{eqnarray}
This Dirac representation of $H({\bf q})$ prompts us to investigate the topological order of this model by means of a degree of map method\cite{vonGersdorff21_unification}. This method introduces a universal topological invariant to describe the topological order of Dirac models in any dimension and symmetry class\cite{vonGersdorff21_unification}, constructed from the unit vector
\begin{eqnarray}
{\bf n}({\bf q})\equiv{\bf d}/|{\bf d}|=(\tilde{n}_{3},\tilde{n}_{4},\tilde{n}_{5}),
\label{nk_unit_vector}
\end{eqnarray}
In 2D systems, the universal topological invariant counts the number of times the BZ torus $T^{2}$ wraps around the target sphere $S^{2}$ formed by the unit vector ${\bf n}({\bf q})$ in Eq.~(\ref{nk_unit_vector}), which is referred to as the wrapping number or degree of the map ${\rm deg}[{\bf n}]$ or the Kronecker index of vector field ${\bf{n}}({\bf{q}})$, and which takes the following cyclic derivative form\cite{vonGersdorff21_unification} 
\begin{eqnarray}
{\rm deg}[{\bf n}]&=&\frac{1}{V_{2}}\int d^{2}{\bf q}\,\varepsilon^{abc}\tilde{n}_{a}\partial_{x}\tilde{n}_{b}\partial_{y}\tilde{n}_{c}
\nonumber \\
&=&\frac{1}{V_{2}}\int d^{2}{\bf q}\,\varepsilon^{abc}\frac{1}{d^{3}}d_{a}\partial_{x}d_{b}\partial_{y}d_{c},
\nonumber \\
&=&\frac{1}{V_{2}}\int d^{2}{\bf q}\,J_{\bf q},
\label{wrapping_number}
\end{eqnarray}
where---following the expansion of Dirac Hamiltonian in Eq.~(\ref{Hk4by4_Gamma345})---$\left\{a,b,c\right\}=\left\{3,4,5\right\}$, $\partial_{\mu}=\partial/\partial q_{\mu}$ is the corresponding momentum derivative, and $V_{2}=4\pi$ is the area of the unit sphere. The integrand $J_{\bf q}\equiv\varepsilon^{abc}\tilde{n}_{a}\partial_{x}\tilde{n}_{b}\partial_{y}\tilde{n}_{c}
=\varepsilon^{abc}d_{a}\partial_{x}d_{b}\partial_{y}d_{c}/d^{3}$ represents the Jacobian of the map $T^{2}\rightarrow S^{2}$, which is also known as the spin Berry curvature of the system\cite{Chen23_spin_Chern}.

Explicitly near the ${\bf K}$ and ${\bf K'}$ points, we expand the momentum by
\begin{eqnarray}
{\bf q}={\bf K}^{(\prime)}+{\bf k}.
\end{eqnarray}
The ${\bf d}=(d_{1},d_{2},d_{3},d_{4},d_{5})$ vector is given in terms of small momentum ${\bf k}$ by
\begin{eqnarray}
&&{\bf K}:\;\;\;{\bf d}=(0,0,v_{F}k_{x},v_{F}k_{y},\lambda_{I}),
\nonumber \\
&&{\bf K}':\;\;\;{\bf d}=(0,0,-v_{F}k_{x},v_{F}k_{y},-\lambda_{I}).
\label{KKp_d1d2d3d4d5}
\end{eqnarray}
where $v_{F}=\sqrt{3}t/2\sim 10^6$m/s is the Fermi velocity of graphene,
which yields the same Jacobian near ${\bf K}$ and ${\bf K}'$ points
\begin{eqnarray}
J_{\bf k}^{\bf K}=J_{\bf k}^{\bf K'}=\frac{\lambda_{I} v_{F}^{2}}{\left[\lambda_{I}^{2}+v_{F}^{2}k^{2}\right]^{3/2}},
\label{KKp_Jacobian}
\end{eqnarray}
and they can be analytically integrated to give 
\begin{eqnarray}
&&{\rm deg}[{\bf n}]^{\bf K}=\frac{1}{4\pi}\int_{0}^{2\pi}d\phi\int_{0}^{\infty}kdk\,J_{\bf k}^{\bf K}
\nonumber \\
&&=\frac{1}{2}={\rm deg}[{\bf n}]^{\bf K'},
\end{eqnarray}
suggesting that the system has integer wrapping number ${\rm deg}[{\bf n}]={\rm deg}[{\bf n}]^{\bf K}+{\rm deg}[{\bf n}]^{\bf K'}=1$. Alternatively, one can numerically integrate the full momentum profile of the Jacobian shown in Fig.~\ref{fig:graphene_Ek_Jk}(b) over the entire BZ, which also gives ${\rm deg}[{\bf n}]=1$, confirming a nontrivial topology of graphene induced by ISOC---a key ingredient for the quantum spin Hall effect (QSHE) in the Kane-Mele model\cite{Kane05,Kane05_2}. 



\subsection{Quantum geometry induced by ISOC}

We proceed by computing the quantum metric of graphene in the presence of ISOC, which for the corresponding Hamiltonian in the Dirac form, Eq.~(\ref{Hk4by4_Gamma345}), can be written as follows \cite{vonGersdorff21_metric_curvature}
\begin{eqnarray}
g_{\mu\nu}=\frac{1}{2d^{2}}\left\{\sum_{i=3}^{5}\partial_{\mu}d_{i}\partial_{\nu}d_{i}
-\partial_{\mu}d\partial_{\nu}d\right\}.
\label{gmunu_generic_Dirac}
\end{eqnarray}
Near ${\bf K}$ and ${\bf K}'$ points, using the expression in Eq.~(\ref{KKp_d1d2d3d4d5}), the components of quantum metric are
\begin{eqnarray}
&&g_{xx}=\frac{1}{2d^{4}}\left[v_{F}^{2}\lambda_{I}^{2}+v_{F}^{4}k_{y}^{2}\right],
\nonumber \\
&&g_{yy}=\frac{1}{2d^{4}}\left[v_{F}^{2}\lambda_{I}^{2}+v_{F}^{4}k_{x}^{2}\right],
\nonumber \\
&&g_{xy}=-\frac{v_{F}^{4}k_{x}k_{y}}{2d^{4}}.
\label{graphene_gmunu_low_energy}
\end{eqnarray}
In terms of the quantum metric the corresponding Jacobian of the map $T^2 \rightarrow S^2$ reads
\begin{eqnarray}
|J_{\bf k}^{\bf K}|=|J_{\bf k}^{\bf K^{\prime}}|=\left(\frac{8}{N}\right)^{\frac{D}{2}}\sqrt{\det g}.
\end{eqnarray}
This relation---known as the metric-curvature correspondence---is ubiquitously satisfied for any Dirac model\cite{vonGersdorff21_metric_curvature},
and therefore plugging for the dimension of Hamiltonian $N=4$ and for the underlying spatial dimension $D=2$ one recovers the result already expressed by Eq.~(\ref{KKp_Jacobian}).


\begin{figure}[ht]
\begin{center}
\includegraphics[clip=true,width=0.99\columnwidth]{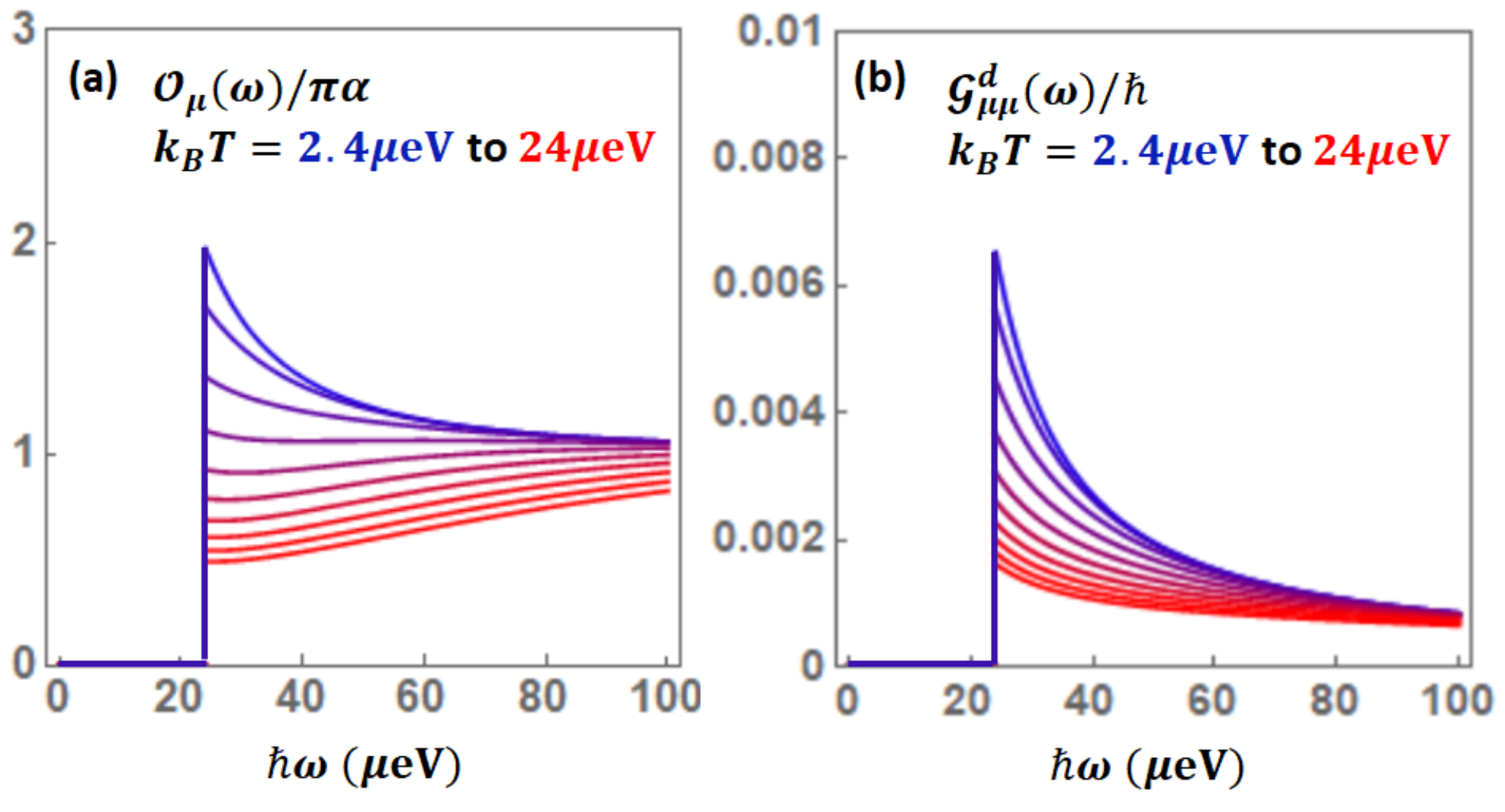}
\caption{(a) The low frequency absorbance ${\cal O}_{\mu}(\omega)$ and (b) fidelity number spectral function ${\cal G}_{\mu\mu}^{d}(\omega)$ at any polarization ${\hat{\boldsymbol\mu}}$ in the presence of ISOC $\lambda_{I}=12\mu$eV and the orbital hopping $t=2.8$\,eV, plotted at different values of temperature (solid lines with different colors) ranging from $k_{B}T=2.4\mu$eV$\approx 0.02$K to $24\mu$eV$\approx 0.2$K.  } 
\label{fig:graphene_ISOC_absorbance_Gw}
\end{center}
\end{figure}

Including contributions from both ${\bf K}$ and ${\bf K'}$ points and using Eq.~(\ref{graphene_gmunu_low_energy}), the low frequency parts of the fidelity number spectral function, Eq.~(\ref{fidelity_number_spec_fn}) and the absorbance, Eq.~(\ref{opacity_wGw}), behave in the same way as for Chern insulators\cite{deSousa23_fidelity_marker}
\begin{eqnarray}
&&{\cal G}_{\mu\mu}^{d}(\omega)=\left[\frac{1}{4\pi\omega}+\frac{\lambda_{I}^{2}}{\pi\hbar^{2}\omega^{3}}\right]
Z(\omega)|_{\omega\geq 2\lambda_{I}/\hbar},
\nonumber \\
&&\frac{{\cal O}_{\mu}(\omega)}{\pi\alpha}=\left[1+\frac{4\lambda_{I}^{2}}{\hbar^{2}\omega^{2}}\right]
Z(\omega)|_{\omega\geq 2\lambda_{I}/\hbar},
\nonumber \\
&&Z(\omega)=\left[f\left(-\frac{\hbar\omega}{2}\right)-f\left(\frac{\hbar\omega}{2}\right)\right],
\end{eqnarray}
which are shown in Fig.~\ref{fig:graphene_ISOC_absorbance_Gw} for a realistic value of ISOC $\lambda_{I}=12\mu$eV$\approx 0.1$K and different values of temperature. We find that the ISOC diminishes the optical absorption at frequency smaller than the gap $\hbar\omega<2\lambda_{I}$, as expected, which for the realistic values lies in the microwave range. At low temperature $k_{B}T\apprle 2\lambda_{I}$, the absorbance at frequency close to the gap $\hbar\omega\apprge 2\lambda_{I}$ is enhanced by the ISOC. Interestingly, at zero temperature $k_{B}T=0$ and at the frequency equal to the gap $\hbar\omega=2\lambda_{I}$, the absorbance ${\cal O}_{\mu}/\pi\alpha=2$ is exactly twice the topologically protected value, and is independent of the polarization of light. We anticipate that these unique features can be used to experimentally quantify the ISOC, provided that the measurement is performed at low enough temperature and the Fermi level is tuned to be inside the gap. If the temperature rises $\hbar\omega\apprge 2\lambda_{I}$, the thermal broadening caused by the Fermi function will reduce the absorbance, as shown in Fig.~\ref{fig:graphene_ISOC_absorbance_Gw}.


Let us provide the graphene's fidelity number at zero temperature by utilizing the full Hamiltonian, Eq.~(\ref{HK_whole_BZ}), with realistic parameters $\lambda_{I}=12\mu$eV for ISOC and $t=2.8$eV for the orbital nearest-neighbor hopping. Direct calculation
gives ${\cal G}_{xx}=1.08$ and ${\cal G}_{yy}=1.28$, signifying a $\sim 20\%$ difference between the absorption of light when polarized along zigzag and armchair directions. Correspondingly, the spread of Wannier function is $\Omega_{I}=2.36\,A_{\rm cell}=2.36\times(\sqrt{3}/2)a_{L}^{2}$, which is slightly larger than the area of unit cell, indicating that the tight-binding Hamiltonian of graphene within the nearest and next-nearest neighbor approximation does provide internally self-consistent results.


\section{Twisted bilayer graphene \label{sec:TBG}}

We now turn to the TBG with a twist angle $\sim 1^{\circ}$ whose intricate phase diagram including superconducting phase\cite{Cao18-2} still attracts 
a lot of attention.  
Interactions that are furnished by the flat bands and specifics of the underlying Moir\'{e} pattern may be presumably responsible for the onset of unconventional superconductivity in TBG. Motivated by the theory that the quantum metric determines the superfluid stiffness\cite{Peotta15,Julku16,Rossi21,Torma21,Xie20} of the flat bands, we compute TBG's fidelity number spectral function. As it comes out, the latter can be detected by the absorbance of light in the millimeter wave range due to the spectral width of the flat bands that spans a range of few meV. This may be a way to verify experimentally whether the average quantum metric in the normal state is truly correlated to the onset of superconductivity in TBG.



\subsection{Minimal two-band model}

We follow the two-band model of Bennett et al\cite{Bennett23} to calculate the absorbance of TBG over a finite range of twist angle. Given the Moir\'{e} period $a_{\rm sc}$, they define the vectors 
\begin{eqnarray}
&&{\bf a}_{1}=a_{\rm sc}{\hat{\bf x}},\;\;\;{\bf a}_{2}=a_{\rm sc}\left(\frac{1}{2}{\hat{\bf x}}+\frac{\sqrt 3}{2}{\hat{\bf y}}\right),
\nonumber \\
&&{\bf a}_{3}=a_{\rm sc}\left(-\frac{1}{2}{\hat{\bf x}}+\frac{\sqrt 3}{2}{\hat{\bf y}}\right),\;\;\;{\bf b}_{1}=a_{\rm sc}\left(\frac{1}{2}{\hat{\bf x}}+\frac{1}{2\sqrt{3}}{\hat{\bf y}}\right),
\nonumber \\
&&{\bf b}_{2}=a_{\rm sc}\left(-\frac{1}{2}{\hat{\bf x}}+\frac{1}{2\sqrt{3}}{\hat{\bf y}}\right),\;\;\;{\bf b}_{3}=-\frac{1}{\sqrt{3}}{\hat{\bf y}}.
\end{eqnarray}
and the following hopping functions 
\begin{eqnarray}
&&f_{1}=\sum_{j=1}^{3}e^{i{\bf k\cdot b}_{j}/\hbar},\;\;\;
f_{2}=\sum_{j=1}^{3}\cos\frac{{\bf k\cdot a}_{j}}{\hbar},
\nonumber \\
&&f_{3}=\sum_{j=1}^{3}e^{-2i{\bf k\cdot b}_{j}/\hbar},
\end{eqnarray}
such that the effective model is described by the $2\times 2$ 
Hamiltonian in the momentum space
\begin{eqnarray}
H^{\rm eff}=H^{\ast}-H^{\rm int}(H^{\Delta})^{-1}{H^{\rm int}}^\dag=
\left(\begin{array}{cc}
d_{0} & d_{-} \\
d_{+} & d_{0}
\end{array}\right).
\label{Heff_TBG}
\end{eqnarray}
To be specific, the subsidiary Hamiltonians $H^I$, where---following the indexing of Ref.~\onlinecite{Bennett23}---$I=\{\ast,\Delta,{\rm int}\}$, have the following generic form
\begin{eqnarray}
H^{I}=
\left(\begin{array}{cc}
t_{0}^{I}+t_{2}^{I}f_{2} & t_{1}^{I}f_{1}+t_{3}^{I}f_{3} \\
t_{1}^{I}f_{1}^{\dag}+t_{3}^{I}f_{3}^{\dag} & t_{0}^{I}+t_{2}^{I}f_{2}
\end{array}\right)
=
\left(\begin{array}{cc}
d_{0}^I & d_{-}^I \\
d_{+}^I & d_{0}^I
\end{array}\right),
\label{TBG_Hast_parametrization}
\end{eqnarray}
and the corresponding hopping parameters 
$\left\{t_{0}^{I},t_{1}^{I},t_{2}^{I},t_{3}^{I}\right\}$
are real, except of $t_{3}^{\rm int}$, which has both real and imaginary part. Moreover, as $H^I$'s and $H^{\rm eff}$ are matrices $2\times 2$ we correspondingly decompose them in terms of the identity matrix $\sigma_0$ and Pauli matrices
$\sigma_\pm=\sigma_1\pm i\sigma_2$ as already obvious from the above expressions.
All ${\bf k}$-dependencies are hidden in the $(d_0^I,d_1^I,d_2^I)\leftrightarrow(d_0^I,d^I_{\pm}=d_1^I\pm i d_2^I)$-coefficients, 
which have the following form for $H^I$
\begin{eqnarray}
&&d_{0}^{I}=t_{0}^I+t_{2}^I\sum_{j=1}^{3}\cos{\bf k\cdot a}_{j},
\nonumber \\
&&d_{1}^{I}=t_{1}^I\sum_{j=1}^{3}\cos{\bf k\cdot b}_{j}
+
{\rm Re}\,t_{3}^I\sum_{j=1}^{3}\cos 2{\bf k\cdot b}_{j}
\nonumber \\
&&\;\;\;\;\;+{\rm Im}\,t_{3}^I\sum_{j=1}^{3}\sin 2{\bf k\cdot b}_{j},
\nonumber \\
&&d_{2}^{I}=-t_{1}^I\sum_{j=1}^{3}\sin{\bf k\cdot b}_{j}+
{\rm Re}\,t_{3}^I\sum_{j=1}^{3}\sin 2{\bf k\cdot b}_{j}
\nonumber \\
&&\;\;\;\;\;-{\rm Im}\,t_{3}^I\sum_{j=1}^{3}\cos 2{\bf k\cdot b}_{j},
\end{eqnarray}
and the corresponding form for $H^{\rm eff}$
\small{\begin{eqnarray}
&&d_{0}=
d_{0}^{\ast}
\nonumber \\
&&-\frac{1}{d_{\Delta}^{2}}\left[d_{0}^{\rm int}d_{0}^{\Delta}d_{0}^{\rm int}-d_{-}^{\rm int}d_{+}^{\Delta}d_{0}^{\rm int}
-d_{0}^{\rm int}d_{-}^{\Delta}d_{+}^{\rm int}+d_{-}^{\rm int}d_{0}^{\Delta}d_{+}^{\rm int}\right],
\nonumber \\
&&d_{-}=d_{1}-id_{2}=(d_{+})^{\dag}
\nonumber \\
&&=d_{1}^{\ast}-id_{2}^{\ast}
-\frac{1}{d_{\Delta}^{2}}\left[2d_{0}^{\rm int}d_{0}^{\Delta}d_{-}^{\rm int}-d_{-}^{\rm int}d_{+}^{\Delta}d_{-}^{\rm int}
-d_{0}^{\rm int}d_{-}^{\Delta}d_{0}^{\rm int}\right],
\nonumber \\
\label{TBG_d0d1d2}
\end{eqnarray}}
where $d_{\Delta}^{2}=(d_{0}^{\Delta})^{2}-(d_{1}^{\Delta})^{2}-(d_{2}^{\Delta})^{2}$.

Although the diagonal $d_{0}$ component of $H^{\rm eff}$ enters the 
eigenenergies
\begin{eqnarray}
\varepsilon_{n}=d_{0}-d,\;\;\;\varepsilon_{m}=d_{0}+d,\;\;\;
d=\sqrt{d_{1}^{2}+d_{2}^{2}},
\end{eqnarray}
it does not enter the corresponding valence-band and conduction-band eigenstates $|n\rangle$ and $|m\rangle$
\begin{eqnarray}
|n\rangle=\frac{1}{\sqrt{2}d}\left(\begin{array}{c}
d \\
d_{1}+id_{2}
\end{array}\right),\;
|m\rangle=\frac{1}{\sqrt{2}d}\left(\begin{array}{c}
-d \\
d_{1}+id_{2}
\end{array}\right).
\label{eq:eigenstatesTBG}
\end{eqnarray}
As the effective model Hamiltonian takes the Dirac form $H^{\rm eff}=\sum_{i=0}^2 d_i\sigma_i$ the quantum metric defined from the lower band eigenstate $|n\rangle$ is according to Eqs.~(\ref{uval_gmunu}) and (\ref{gmunu_generic_Dirac}) given by 
\begin{eqnarray}
&&g_{\mu\nu}=\frac{1}{4d^{2}}\left\{\partial_{\mu}d_{1}\partial_{\nu}d_{1}
+\partial_{\mu}d_{2}\partial_{\nu}d_{2}-\partial_{\mu}d\partial_{\nu}d\right\},
\nonumber \\
&&=\frac{1}{8d^{2}}\left\{\partial_{\mu}d_{+}\partial_{\nu}d_{-}
+\partial_{\mu}d_{-}\partial_{\nu}d_{+}-2\partial_{\mu}d\partial_{\nu}d\right\}.
\label{TBG_gmunu}
\end{eqnarray}
Finally, because the above calculation does not take into account the spin degeneracy, the full quantum metric is twice the above result
\begin{eqnarray}
g_{\mu\nu}\rightarrow g_{\mu\nu}\times 2,
\end{eqnarray}
and so follows the extra factor of two also for the fidelity number spectral function and opacity.

\begin{figure}[ht]
\begin{center}
\includegraphics[clip=true,width=0.99\columnwidth]{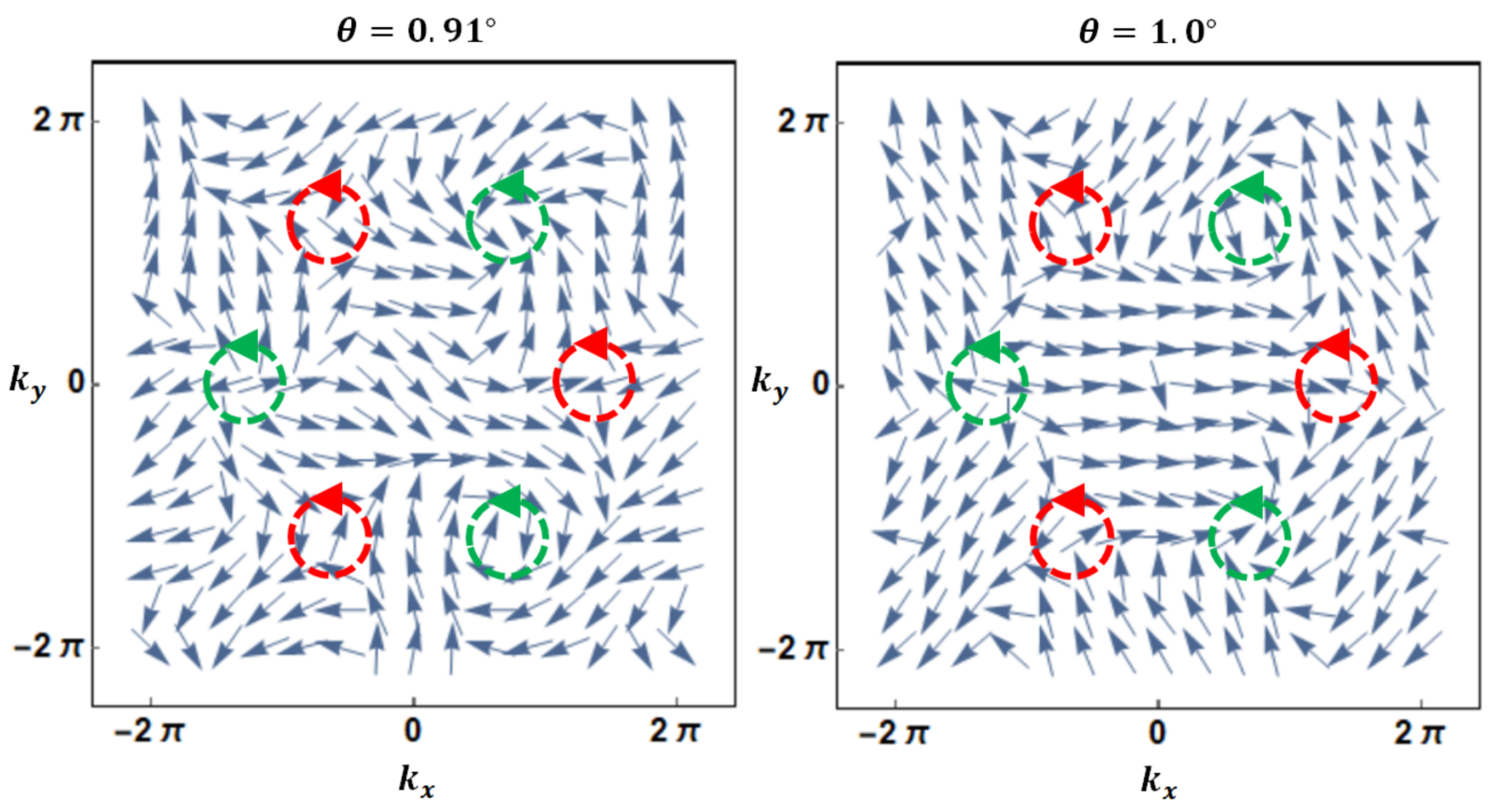}
\caption{The momentum space profile of the unit vector ${\bf n}=(\tilde{n}_{1},\tilde{n}_{2})={\bf d}/|{\bf d}|$ that characterizes the momentum-dependence of the effective Hamiltonian of TBG in Eq.~(\ref{Heff_TBG}) plotted as a vector field at two twist angles $\theta=0.91^{\circ}$ and $1.0^{\circ}$. As going along a loop circulating the nodal points counterclockwisely, one sees that ${\bf K}$ (red circles) and ${\bf K'}$ points (green circles) have opposite winding numbers of the vector field. } 
\label{fig:TBG_winding_number}
\end{center}
\end{figure}

\subsection{Topological charges and metric-curvature correspondence of TBG}

Now we demonstrate that the topological charges at the nodal points of TBG are likewise to
those in the pristine graphene, see Sec.~\ref{sec:graphene_ISOC}, if ISOC is absent.
Taking the valence band state $|n\rangle$ given by Eq.~(\ref{eq:eigenstatesTBG})
and choosing an arbitrary closed contour circulating any of the nodal points, such that
$\phi$ denotes the corresponding azimuthal angle, the quantum metric $g_{\phi\phi}$ and the valence band Berry connection $\langle n|i\partial_{\phi}|n\rangle$ satisfies the same metric-curvature correspondence as in the pristine graphene\cite{vonGersdorff21_metric_curvature}
\begin{eqnarray}
&&g_{\phi\phi}=\langle\partial_{\phi}n|\partial_{\phi}n\rangle-\langle\partial_{\phi}n|n\rangle
\langle n|\partial_{\phi}n\rangle=|\langle n|i\partial_{\phi}|n\rangle|^{2}
\nonumber \\
&&=\left|\frac{1}{2}\varepsilon^{ab}\tilde{n}_{a}\partial_{\phi}\tilde{n}_{b}\right|^{2},
\label{TBG_metric_curvature}
\end{eqnarray}
where ${\bf n}=(\tilde{n}_{1},\tilde{n}_{2})=(d_{1},d_{2})/d$ is the unit vector, 
The topological charge of each nodal point is calculated from the integration of Berry connection over the closed contour 
\begin{eqnarray}
&&{\cal C}=\oint\frac{d\phi}{2\pi}\langle n|i\partial_{\phi}|n\rangle=-\frac{1}{2}\oint\frac{d\phi}{2\pi}\varepsilon^{ab}\tilde{n}_{a}\partial_{\phi}\tilde{n}_{b}
\nonumber \\
&&=\pm\frac{1}{2}{\rm deg}\left[{\bf n}\right]=\pm\frac{1}{2}.
\label{TBG_topological_charge}
\end{eqnarray}
which is equivalently the winding number of the vector field ${\bf n}=(\tilde{n}_{1},\tilde{n}_{2})$ along the trajectory. The topological charges are of opposite signs at the ${\bf K}$ and ${\bf K}'$ points at any twist angle (for which the Hamiltonian $H^{\rm eff}$ makes a meaningful approximation), as can be seen from the winding of the ${\bf n}$ field shown in Fig.~\ref{fig:TBG_winding_number} using, for example, the twist angle $0.91^{\circ}$ and $1.0^{\circ}$. This analysis indicates that TBG remains as much topological (or Dirac) semimetal as the monolayer graphene provided the ISOC is disregarded.

\begin{figure*}[ht]
\begin{center}
\includegraphics[clip=true,width=1.95\columnwidth]{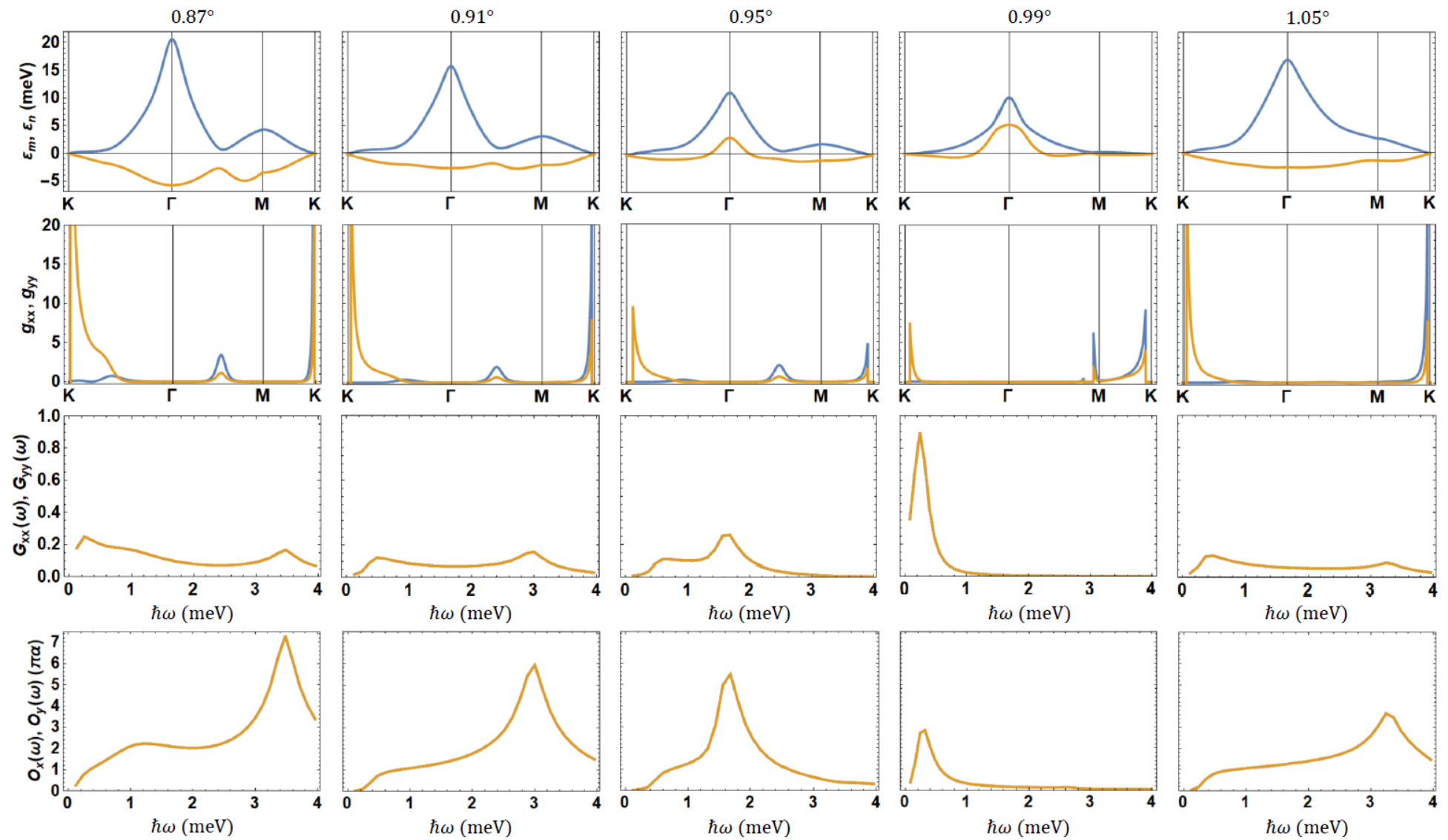}
\caption{Numerical results for TBG at five selected twist angles $\theta=0.87^{\circ},0.91^{\circ}, 0.95^{\circ}, 0.99^{\circ},1.05^{\circ}$, where from the top to the bottom we present the band structure $(\varepsilon_{n},\varepsilon_{m})$ and the quantum metric $(g_{xx},g_{yy})$ along high-symmetry lines, followed by the fidelity-number spectral function $({\cal G}_{xx}(\omega),{\cal G}_{yy}(\omega))$ and the absorbance under polarized light $({\cal O}_{x}(\omega),{\cal O}_{y}(\omega))$ as
functions of energy $\hbar\omega$.} 
\label{fig:TBG_figure}
\end{center}
\end{figure*}


\subsection{Quantum geometry and low frequency opacity of TBG}

The numerical results for the band structure $\left\{\varepsilon_{n},\varepsilon_{m}\right\}$, quantum metric $g_{\mu\nu}$, fidelity number spectral function ${\cal G}_{\mu\nu}(\omega)$, and opacity ${\cal O}_{\mu}(\omega)$ resulting from the tight-binding TBG model of Bennett et al\cite{Bennett23} are shown in Fig.~\ref{fig:TBG_figure} for a few selected twist angles. 
The ${\cal G}_{\mu\nu}(\omega)$ and ${\cal O}_{\mu}(\omega)$ are calculated numerically from Eq.~(\ref{sigmaw_gmumu}) by approximating the $\delta$-function in the optical transition by a Lorentzian shape
\begin{eqnarray}
\delta\left(\omega+\frac{\varepsilon_{n}}{\hbar}-\frac{\varepsilon_{m}}{\hbar}\right)
\simeq\frac{\eta/\pi}{\left(\omega+\frac{\varepsilon_{n}}{\hbar}-\frac{\varepsilon_{m}}{\hbar}\right)^2+\eta^{2}},
\end{eqnarray}
where $\eta=0.1$meV is a phenomenological broadening parameter simulating effects of correlations, and we will restrict our discussion to zero temperature. The band structure always exhibits Dirac-like dispersion around the ${\bf K}$ and ${\bf K}'$ points, but the chemical potential lies at some $0.1\sim 0.5$meV above the Dirac point according to the model of Bennett et al\cite{Bennett23}. Shall the chemical potential lies exactly at the Dirac points, the quantum metric 
$\left\{g_{xx},g_{yy}\right\}$ would diverge at the Dirac points, and consequently also
the fidelity number spectral function $\left\{{\cal G}_{xx}(\omega),{\cal G}_{xx}(\omega)\right\}\sim 1/\omega$ for low frequencies\cite{deSousa23_fidelity_marker}, yielding, however, a frequency-independent opacity\cite{deSousa23_graphene_opacity} 
for low values of $\omega$'s. 
But because of the finite chemical potential $\mu$, the quantum metric in the range from the Dirac point to $\mu$ nullifies since both bands $\left\{\varepsilon_{n},\varepsilon_{m}\right\}$ are either occupied ($\mu>0$) or empty ($\mu<0$) and hence the divergence of $\left\{g_{xx},g_{yy}\right\}$ at the ${\bf K}$ and ${\bf K}'$ points and that of $\left\{{\cal G}_{xx}(\omega),{\cal G}_{xx}(\omega)\right\}$ for $\omega\rightarrow 0$ get removed. Consequently the opacity at $\omega\rightarrow 0$ is suppressed, a phenomenon similar to fluorinated graphene where the chemical potential is shifted from the Dirac point\cite{Nair10,deSousa23_graphene_opacity}.

Naively, the TBG should behave like two independent monolayer graphene sheets, and therefore one expects its absorbance to be roughly twice that of monolayer graphene by counting the optical absorption power
\begin{eqnarray}
\overline{{\cal O}}(\omega)=\pi\alpha+(1-\pi\alpha)\times\pi\alpha\approx 2\pi\alpha.
\end{eqnarray}
We found that this naive counting is true only for frequencies larger than chemical potential of the order of $\hbar\omega\apprge 0.5$meV at most twist angles, but at $\hbar\omega\apprle 0.5$meV the absorbance is suppressed by finite chemical potential as explained above. Moreover, in the magic angle region $0.96^{\circ}\leq\theta\leq 1.02^{\circ}$ 
this model predicts particularly flat valence and conduction bands whose shapes are very similar, but just separated by a very small energy. Consequently, the absorbance has a very interesting behavior: In this region (see the $\theta=0.99^{\circ}$ data in Fig.~\ref{fig:TBG_figure} as an example), both the fidelity number spectral function ${\cal G}_{xx}(\omega)\approx{\cal G}_{yy}(\omega)$ and opacity ${\cal O}_{\mu}(\omega)$ peak at very small frequency $\hbar\omega\approx 0.3$meV, and at higher frequency they are practically vanishing. These low frequency peaks originate from the fact that all the momenta in a large region of the BZ absorb light equivocally at almost the same frequency (the energy difference between the two flat bands). This peak of ${\cal G}_{\mu\mu}(\omega)$ at $\hbar\omega\approx 0.3$meV also implies that the quantum metric is mainly distributed in the flat band region around $\Gamma-M$ (see the $\left\{\varepsilon_{n},\varepsilon_{m}\right\}$ plot for $\theta=0.99^{\circ}$). Thus our result seems to suggest the coincidence between the increased quantum metric and the formation of flat bands at the magic angle, hence supporting the quantum metric theory of the superfluid density\cite{Peotta15,Julku16,Liang17,HerzogArbeitman22,Torma22,Iskin23}.

\begin{figure}[ht]
\begin{center}
\includegraphics[clip=true,width=0.99\columnwidth]{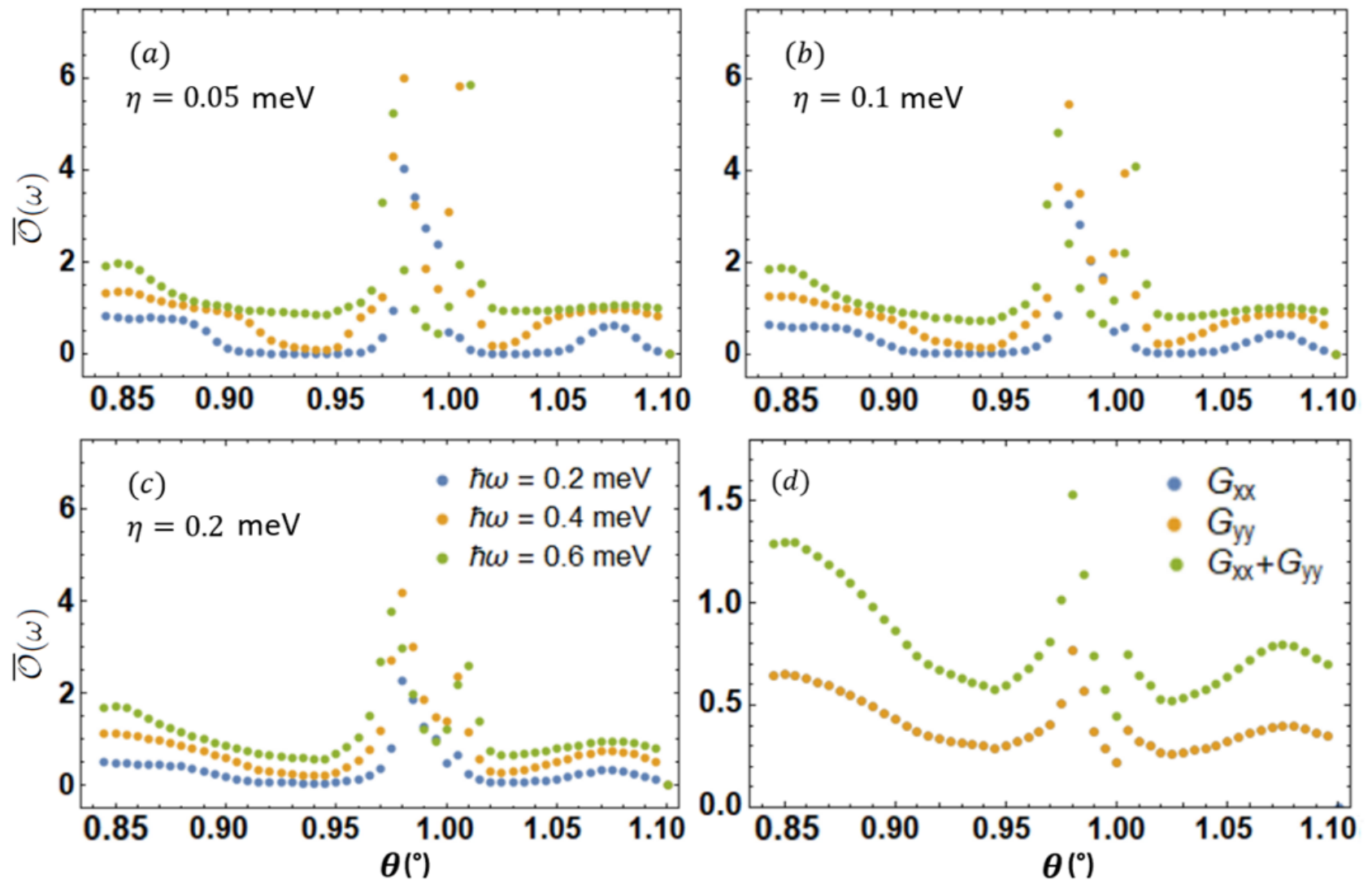}
\caption{Opacity of TBG under unpolarized light $\overline{{\cal O}}(\omega)$ at zero temperature as a function of twist angle $\theta$ plotted at three different frequencies 
$\hbar\omega=(0.2,0.4,0.6)$meV, and using three different values of the artificial broadening (a) $\eta=0.05$meV, (b) $\eta=0.1$meV, (c) $\eta=0.2$meV that phenomenologically simulate different strengths of correlations. (d) The fidelity number $\left\{{\cal G}_{xx},{\cal G}_{yy},{\cal G}_{xx}+{\cal G}_{yy}\right\}$ as a function of twist angle $\theta$.} 
\label{fig:TBG_opacity_versus_theta}
\end{center}
\end{figure}

Our quantum metric analysis of the flat band properties of TBG implies a straightforward verification based on a simple optical experiment: If the absorbance in the range $\hbar\omega\sim 0.1$meV is significantly enhanced for the twist angles in the magic angle region, or in other words, if the TBG would become substantially dark for the light in the millimeter wave-lengths as shown in Fig.~\ref{fig:TBG_opacity_versus_theta} (a), (b), and (c) simulated for different values of $\eta$, this would indirectly signify the formation of flat bands. Moreover, dividing the absorbance by frequency and then integrating such ratio over $\omega$ gives the trace of the fidelity number according to Eq.~(\ref{OmegaI_opacity}), hence giving an estimation to the total quantum metric carried by the flat bands. In terms of the effective tight-binding model we estimated this quantity and provided its course as a function of twist angle in Fig.~\ref{fig:TBG_opacity_versus_theta} (d). The latter shows that fidelity numbers ${\cal G}_{xx}$ and ${\cal G}_{yy}$ peak around $\theta\approx 0.98^{\circ}$. It is worth to emphasize that the fidelity number is finite only because the chemical potential cuts off the divergent quantum metric, i.e., the spread of Wannier function $\Omega_{I}=A_{\rm cell}({\cal G}_{xx}+{\cal G}_{yy})$ in fact diverges in this model. This is owing to the artifact that this tight-binding model does not take into account the ISOC at $\mu$eV that is necessary to make the spread finite, as elaborated at the end of Sec.~\ref{sec:graphene_ISOC} for the monolayer graphene.

\section{Transition metal dichalcogenides \label{sec:TMD}}

\subsection{Tight-binding model of TMDs}

Finally, we turn to hexagonal TMDs, $1H-MX_{2}$, where transition metal $M$ = Mo, W; and chalcogen $X$ = S, Se, Te and use a generic tight-binding Hamiltonian of Ref.~\onlinecite{Liu13}. The latter consists of six-band model spanned 
by three transition metal (Mo,W) $d$ orbitals and their two spins projections
\begin{eqnarray}
|\psi\rangle=(d_{z^{2}}\uparrow,d_{xy}\uparrow,d_{x^{2}-y^{2}}\uparrow,
d_{z^{2}}\downarrow,|d_{xy}\downarrow,d_{x^{2}-y^{2}}\downarrow).
\nonumber \\
\end{eqnarray}
The orbital hoppings constitute a $3\times 3$ Hamiltonian for each spin sector 
\begin{eqnarray}
H_{0}({\bf k})=\left(\begin{array}{ccc}
V_{0} & V_{1} & V_{2} \\
V_{1}^{\ast} & V_{11} & V_{12} \\
V_{2}^{\ast} & V_{12}^{\ast} & V_{22}
\end{array}\right)
\end{eqnarray}
where the detailed ${\bf k}$-dependence of $V_{0}$---$V_{22}$ and the set of underlying parameters for each TMD compound can be found in Ref.~\onlinecite{Liu13}. Including the spin-orbit coupling (SOC), the full $6\times 6$ Hamiltonian takes the form
\begin{eqnarray}
&&H({\bf k})=\left(\begin{array}{cc}
H_{0}({\bf k})+\frac{\lambda}{2}L_{z} & 0 \\
0 & H_{0}({\bf k})-\frac{\lambda}{2}L_{z}
\end{array}\right),
\nonumber \\
&&L_{z}=\left(\begin{array}{ccc}
0 & 0 & 0 \\
0 & 0 & 2i \\
0 & -2i & 0
\end{array}\right),
\end{eqnarray} 
where $\lambda$ is the strength of the SOC.

\begin{figure*}[ht]
\begin{center}
\includegraphics[clip=true,width=1.99\columnwidth]{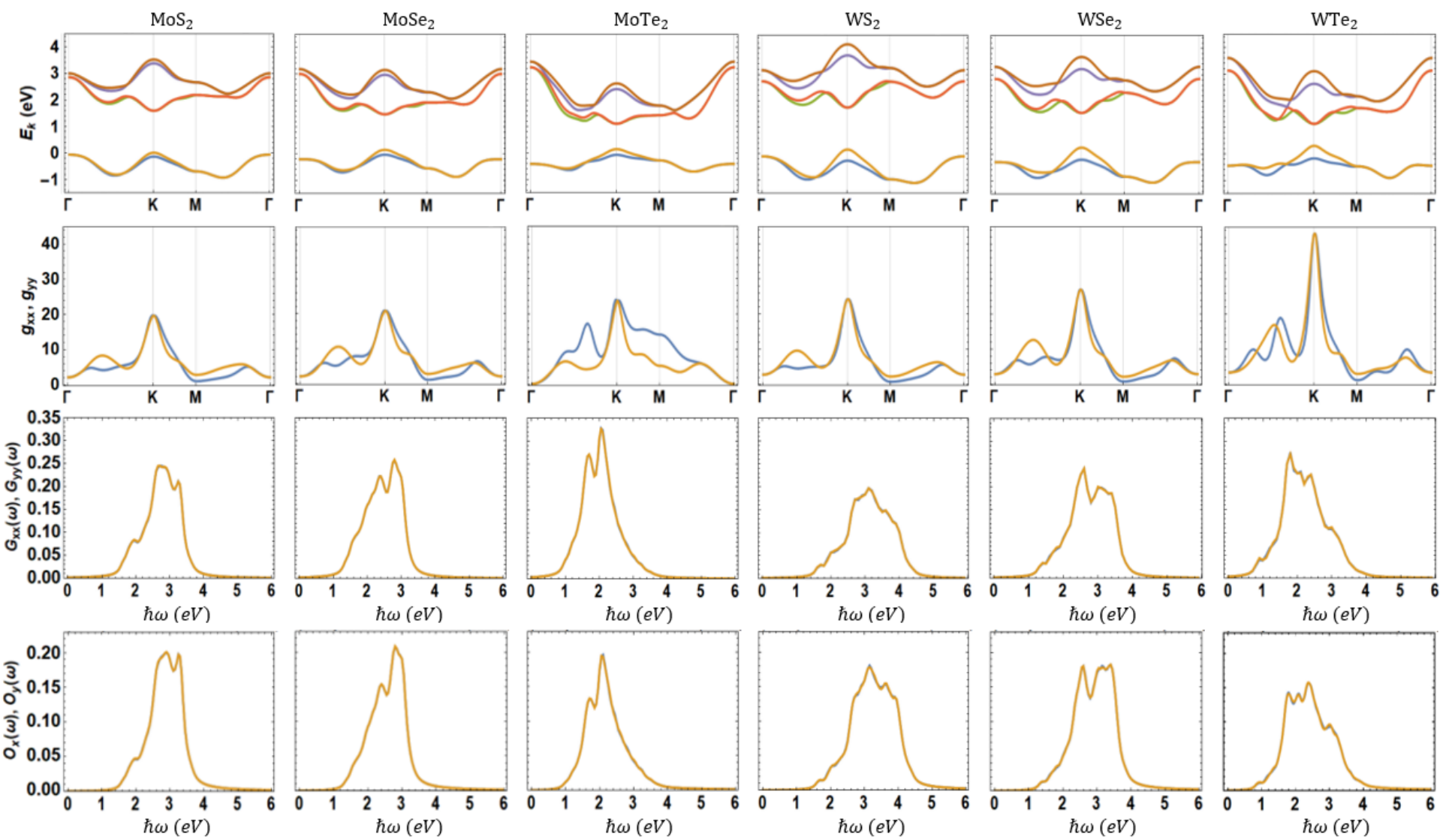}
\caption{Model calculated band structure $(\varepsilon_{n},\varepsilon_{m})$ and 
quantum metric $(g_{xx},g_{yy})$ displayed along the high-symmetry lines in the BZ, followed by fidelity-number spectral function $({\cal G}_{xx}(\omega),{\cal G}_{yy}(\omega))$, and the absorbance $({\cal O}_{x}(\omega),{\cal O}_{y}(\omega))$ under polarized light as functions of energy $\hbar\omega$ for the six TMD materials: MoS$_{2}$, MoSe$_{2}$, MoTe$_{2}$, WS$_{2}$, WSe$_{2}$, and WTe$_{2}$. } 
\label{fig:TMD_figure}
\end{center}
\end{figure*}

Numerical results for the six TMD materials described by this tight-binding model are shown in Fig.~\ref{fig:TMD_figure}, simulated with artificial broadening $\eta=0.1$eV. These materials have a typical semiconducting gap $\sim$ 1eV to $2$eV with one pair of spin-split valence bands and two pairs of spin-split conduction bands, indicating their optical responses fall into the infrared up to visible light range. The quantum metric is found to peak at the ${\bf K}$ and ${\bf K}'$ points, but it does not show a sharp divergence like in pristine graphene, therefore the TMDs fidelity numbers ${\cal G}_{\mu\nu}$'s and their spreads $\Omega_{I}$'s are both finite. 
Calculating the spread directly from either the momentum integration in Eq.~(\ref{Gmunu_definition}) or the frequency integration in Eq.~(\ref{OmegaI_opacity}) yields the $\Omega_{I}$'s summarized in Table \ref{tab:TMD_data}. 
The ratio $\Omega_{I}/A_{\rm cell}={\rm Tr}\,{\cal G}_{\mu\nu}$ of these materials falls in the range of $0.6\sim 0.8$, indicating fairly localized Wannier functions. Finally, we remark that the absorbance $\overline{\cal O}(\omega)$ obtained from this tight-binding model yields an excellent agreement with the experimental data of free standing samples under unpolarized light at low frequencies $\hbar\omega\apprle 3$eV\cite{Li14,Li17}. In other words, the minimal model employing three $d$-orbitals of the transition metal captures at low frequencies the essential optical, as well as, the quantum metric properties very accurately.

\begin{table}[ht]
  \begin{center}
    \caption{Area of the unit cell $A_{\rm cell}$, spread of the valence band Wannier functions $\Omega_{I}$, and the dimensionless ratio $\Omega_{I}/A_{\rm cell}={\rm Tr}{\cal G}_{\mu\nu}$ calculated from the tight-binding model of Ref.~\onlinecite{Liu13}. }
    \label{tab:TMD_data}
    \begin{tabular}{ c c c c c }
    \hline
    {\rm Mat}. & $A_{\rm cell}(\AA^{2})$ & $\Omega_{I}(\AA^{2})$ & $\Omega_{I}/A_{\rm cell}={\rm Tr}{\cal G}_{\mu\nu}$ \\ \hline
    MoS$_{2}$ & 26.44 & 16.95 & 0.641  \\ 
    WS$_{2}$ & 26.45 & 17.22 & 0.651  \\ 
    MoSe$_{2}$ & 28.74 & 19.57 & 0.681 \\ 
    WSe$_{2}$ & 28.72 & 20.36 & 0.709  \\ 
    MoTe$_{2}$ & 32.87 & 24.42 & 0.743 \\ 
    WTe$_{2}$ & 32.93 & 26.31 & 0.799  \\
    \hline
  \end{tabular}
  \end{center}
\end{table}

\subsection{Absorbance of WS$_{2}$ on fused silica \label{sec:WS2_silica}}

In this section, we use the experimental data and demonstrate several realistic issues on extracting the spread $\Omega_{I}$ from the absorbance measurements, including the effect of substrate, presence of higher energy bands, and also of excitons. We synthesized centimeter-scale WS$_{2}$ polycrystalline samples on fused silica (University WAFER, Fused Silica, 500 um thickness) via chemical vapor deposition (CVD)\cite{Stand22}. Since the fused silica wafer exhibits more than $99\%$ transparency and has a band gap larger than 9eV, it allows us to measure the absorbance of the WS$_{2}$ deposited on it up to very high frequencies, and perform the experiment at room temperature. To measure the absorbance, a collimator featuring a circular aperture with a diameter of 2.7mm positioned in the monolayer region of WS$_{2}$ sample was employed, where the monolayer region is identified by Raman spectroscopy\cite{Zeng15,Barbosa20}.

\begin{figure}[ht]
\begin{center}
\includegraphics[clip=true,width=0.99\columnwidth]{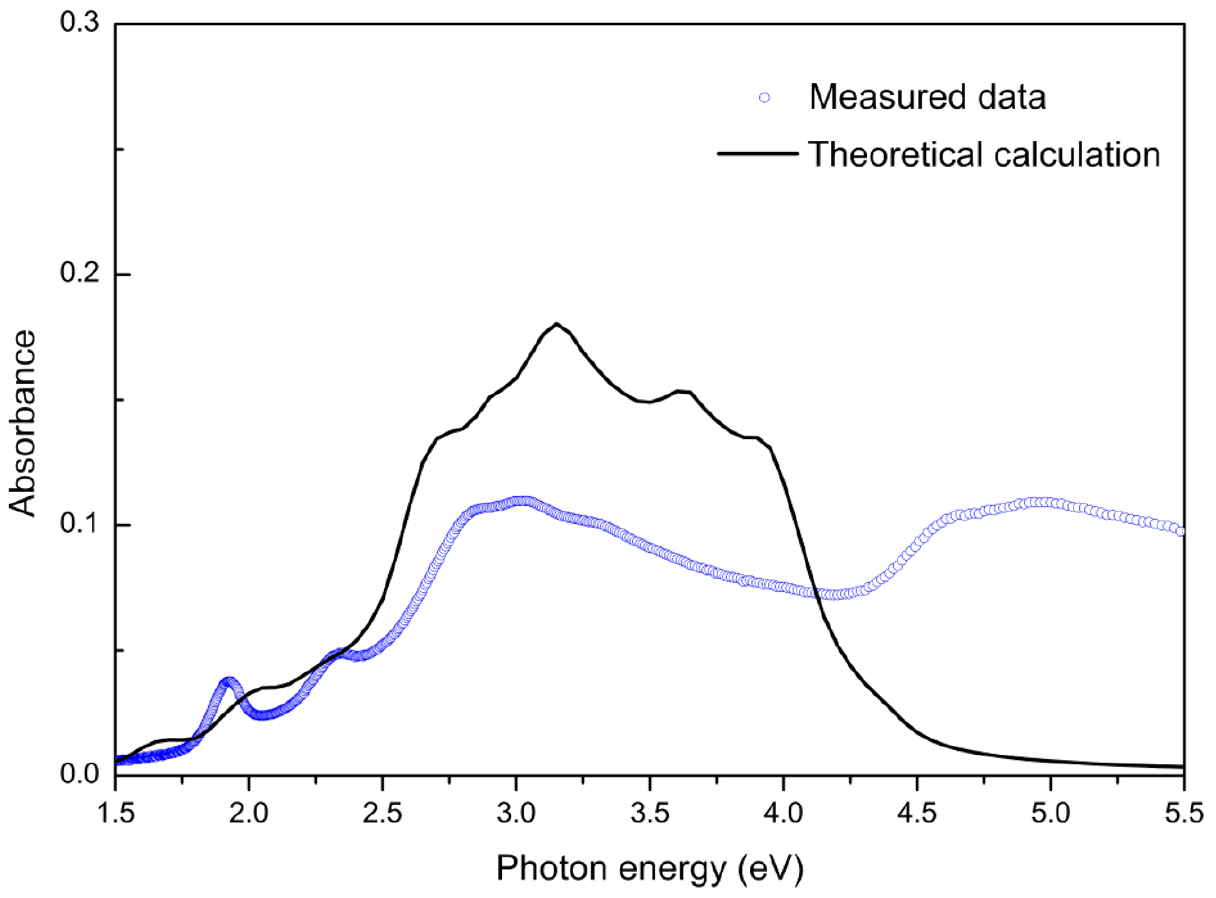}
\caption{Absorbance of WS$_{2}$ measured at room temperature on a fused silica, blue circles, contrasted with theoretical calculation at zero temperature (black line) based on three-orbital model as already shown in Fig.~\ref{fig:TMD_figure}.} 
\label{fig:WS2_experiment}
\end{center}
\end{figure}

The experimental data of the absorbance extracted from the transmission measured by a Lambda 950 UV-Vis-NIR spectrophotometer are shown in Fig.~\ref{fig:WS2_experiment} as blue circles. Comparing with the theoretical results coming from the low-energy tight-binding model described above and displayed in Fig.~\ref{fig:TMD_figure} and Fig.~\ref{fig:WS2_experiment} (black solid line) for a free standing WS$_{2}$, the following differences become apparent. 

First, our experimental data agrees well with the theoretical curve up to about $\hbar\omega\apprle$2.5eV, meaning that the tight-binding model employing just three $d$-orbitals of W captures the absorbance at low frequency range very accurately. However, the experimental absorbance in Fig.~\ref{fig:WS2_experiment} becomes smaller than the theoretical one for higher frequency range $2.5$eV$\apprle \hbar\omega\apprle 4.0$eV, indicating that interfacing with the fused silica substrate potentially reduces the spread of Wannier functions originating from the W orbitals. At even higher frequencies $\hbar\omega\apprge 4$eV, the model-calculated absorbance diminishes, while still a nonzero absorbance is detected experimentally, meaning that the $p$-orbitals of S start to absorb light and contribute to the spread of valence-band Wannier orbitals. As chalcogen orbitals are not included in the minimal tight-binding model of Ref.~\onlinecite{Liu13}, one needs to resort to a more elaborate tight-binding description or first-principle calculations to capture these absorption at higher frequency, which should be addressed elsewhere.

Another issue that can not be avoided in experiment is the formation of excitons, which are responsible for the peaks around 1.9eV and 2.3eV in Fig.~\ref{fig:WS2_experiment}. These excitons do not contribute to the spread $\Omega_{I}$ for the following reasons: (1) They are bosonic excitations, and hence should only occupy a small part of the BZ. Moreover, the notion of Wannier states for excitons is rather ambiguous. (2) They are only present when the light excites the sample, and therefore they are not a genuinely linked to the equilibrium ground state properties of the Fermi sea but rather to light-matter interacting system. This leads us to conclude that shall one intend to extract the spread $\Omega_{I}$ from an experimental absorbance data $\overline{\cal O}(\omega)$ using Eq.~(\ref{OmegaI_opacity}), the exciton absorption peaks should be meaningfully excluded.

\section{Conclusions}

In summary, by employing the perspectives of quantum metric, we propose an optical absorption scheme to estimate the gauge-invariant part of the spread $\Omega_{I}$ of the valence band Wannier orbitals in semiconductors and insulators. The theory we put forward is based on the quantum geometrical origin of the valence band spread, and the fact that the quantum metric of the associated states is formally equivalently to the matrix elements of optical transition. In 3D systems, we reveal that the spread of valence Wannier states can be extracted from the frequency-integral of the imaginary part of the dielectric function multiplied by the volume of unit cell. Such quantity also represents the average distance on the BZ manifold when endowing the latter with a nontrivial (curved) quantum metric, of which we call the fidelity number. Using the experimental data of the dielectric function of Si, Ge, and Bi$_{2}$Te$_{3}$, we extract the absolute scale of $\Omega_{I}$, we interpret the ratio $\Omega_{I}^{3/2}/V_{\rm cell}$ as a figure of merit for estimating the insulating characteristics of these materials, demonstrating the ubiquity of our proposal.

For 2D materials, we propose that the spread divided by unit cell area $\Omega_{I}/A_{\rm cell}$ can be measured by the optical absorbance divided by frequency and then integrated over the frequencies. Applying this method to graphene reveals the importance of ISOC in obtaining a finite spread, as otherwise the quantum metric diverges at Dirac points. 
In addition, the absorbance of graphene at microwave range measured in the sub-Kelvin region can be used to directly estimate the magnitude of ISOC. We further use a recently proposed tight-binding model for structurally-relaxed TBG and examine its quantum geometric and optical-absorption properties in a wide range of twist angle. Based on our analysis we predict an abrupt increase of the absorbance at low frequencies $\hbar\omega\apprle 0.1$meV, which can serve as a feasible test to detect the formation of flat bands and the distribution of quantum metric therein. This may help to verify experimentally whether the unconventional superconductivity in TBG is potentially related to the quantum metric. Finally, applying our method to hexagonal TMD materials yields a low frequency absorbance in an excellent agreement with the experiments, and suggesting that the ratio $\Omega_{I}/A_{\rm cell}$ contributed from the (Mo,W) orbitals is about $0.6\sim 0.8$. Furthermore, comparing our theory with the experimental results of WS$_{2}$ deposited on fused silica reveals that the substrate can greatly reduce $\Omega_{I}$ contributed from a free standing model solely due to W orbitals.

It is worth to emphasize that our quantum metric formalism considers Fermi sea free of any bound electron-hole pairs, therefore when dealing with experimental data the contribution from the excitonic peaks shall be carefully disregarded when calculating $\Omega_{I}$ by integration of the optical absorption data over the frequencies. In addition, although the higher energy absorbance due to the (S,Se,Te) orbitals were not captured by the simple tight-binding model of TMDs, we suggest that it may be captured by more sophisticated first principle calculations. These results indicate that the proposed experimental protocol for extracting $\Omega_{I}$ can be ubiquitously applied to either free-standing or substrated 2D materials of any chemical composition, and can include any complications in realistic experiments.

\begin{acknowledgments}

We thank E. Pavarini, L. Boeri, I. Komissarov, and P. J. Hirschfeld for fruitful discussions, and D. Bennett for sharing the data for TBG. W. C. acknlwledges the financial support of the productivity in research fellowship
from CNPq. D.K.~acknowledges partial support from the IM\-PULZ project IM-2021-26---SUPERSPIN funded by the Slovak Academy of Sciences.

\end{acknowledgments}

\bibliography{Literatur-2}

\end{document}